\documentclass[pra,twocolumn,amsmath,amssymb]{revtex4}

\usepackage{graphicx}   

\usepackage{dcolumn}

\begin{document}

\title{Controlling interactions between highly-magnetic atoms with Feshbach resonances}

\author{Svetlana Kotochigova}
\address{Department of Physics, Temple University, Philadelphia, Pennsylvania 19122}

\begin{abstract}
This paper reviews current experimental and theoretical progress in the
study of dipolar quantum gases of ground and meta-stable atoms with a
large magnetic moment. We emphasize the anisotropic nature of Feshbach
resonances due to coupling to fast-rotating resonant molecular states in
ultracold $s$-wave collisions between magnetic atoms in external magnetic
fields. The dramatic differences in the distribution of resonances
of magnetic $^7$S$_3$ chromium and magnetic lanthanide atoms with a
submerged 4f shell and non-zero electron angular momentum is analyzed.
We focus on  Dysprosium and Erbium as important experimental advances have
been recently made to cool and create quantum-degenerate gases for these atoms.
Finally, we describe progress in locating resonances in collisions of
meta-stable magnetic atoms in electronic P states with ground-state atoms, where
an interplay between collisional anisotropies and spin-orbit coupling exists.
\end{abstract}

\pacs{03.75.Mn, 03.75.Ss, 05.30.Jp, 34.20.Cf, 34.50.Cx}

\maketitle

\tableofcontents
                                                                                                   
\section{Introduction}

Breakthroughs in the experimental realization of ultracold dipolar quantum gases
of atoms with a large magnetic moment, such as Cr \cite{Pfau2005}, Dy
\cite{Lev2011,Lev2012}, and Er \cite{Ferlaino2012A,Ferlaino2012B} have
opened a new scientific playground for the study of strongly-correlated
atomic systems. These breakthrough are not only limited to ground-state atoms,
but apply also to meta-stable $P$ state systems, where the non-zero orbital angular momentum
contributes to a substantial magnetic moment \cite{Takahashi2013,Gupta2014}.
This new research area is enabled by the long-range and
anisotropic nature of the magnetic dipole-dipole interactions between
magnetic atoms that allows to engineer exotic many-body phases with
control and tunability \cite{Fradkin2009,BLev2009}. Due to its large spin,
dipolar gases of magnetic atoms represent
an excellent environment for exploring the interface between condensed matter and atomic physics,
as recently illustrated by \cite{BLaburthe2013}, where a complex spin
dynamics is observed for doubly occupied sites of an optical lattice
containing Cr atoms. In addition, ultracold samples of magnetic atoms
are proposed for precision measurements of parity nonconservation  and
variation of fundamental constants \cite{Budker1994,Nguyen1997}, as
well as quantum information processing \cite{DereviankoQC}. For example,
robust quantum memory can be created with highly-magnetic atoms  coupled
to a super-conducting stripline \cite{Rabl2006}.

\subsection{Characterization of Feshbach resonance}

In this Review we will explore the concept of controlling the interactions between
ultracold magnetic atoms with magnetic Feshbach resonances. 
The limit of infinitely strong interactions between atoms leads to
strongly-interacting quantum gases. Alternatively, interactions can be 
turned to zero to create an ideal Fermi or Bose gas. 

\begin{figure} 
\includegraphics[scale=0.30,trim=0 0 0 0,clip]{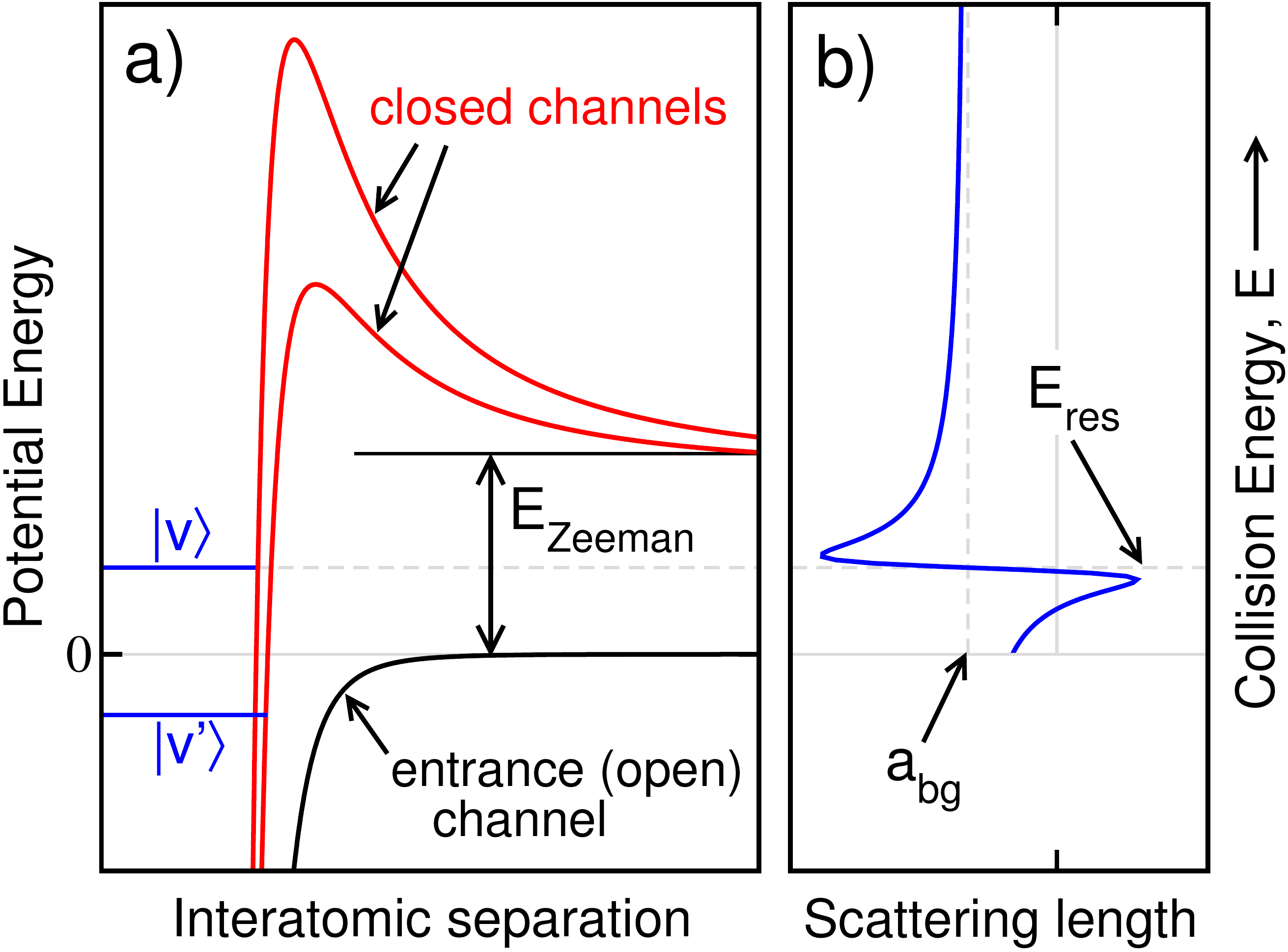}
\caption{a) Schematic of interatomic potentials of magnetic atoms. Scattering starts in the 
entrance and open $s$-wave channel. A Feshbach resonance is due to a bound state $|v\rangle$
of a potential of a closed channel with a dissociation energy above that of the open channel. Here,
 the splitting, $E_{\rm Zeeman}$, is due to the Zeeman interaction
and can be varied with a magnetic field. The vibrational level $|v'\rangle$ of a second closed channel is a stable and weakly-bound molecular level and can become a resonance when $E_{\rm Zeeman}$ is further increased.
Panel b) shows the scattering length as a function of collision energy. A resonance with a distinctive Fano profile \cite{Fano1961} occurs near energy $E_{\rm res}$.}
\label{fig:simpleFR}
\end{figure}

Figure~\ref{fig:simpleFR}a) illustrates the physics of
a Feshbach resonance based on a schematic picture of the interaction
potentials between two atoms.  Typically, in ultracold-atom experiments
a homogeneous magnetic field $B$ is present  and all atoms are prepared
in its energetically-lowest Zeeman sublevel. Two of such atoms form
the entrance or open channel.  Closed channels correspond to pairs of
atoms in energetically-higher Zeeman states.  By changing the magnetic
field strength  the closed channel energy shifts with respect to open
channel energy.

In the course of the collision the open and closed channels couple and  a magnetic Feshbach resonance appears,
when a bound state of a closed channel has an energy near the collision energy of
the open channel. The bound state is resonantly coupled to the continuum. 
As we will discuss below in more detail for collisions of magnetic atoms,
coupling between open and closed channels can only occur due to anisotropic molecular interactions whose strengths depend on the direction along which the atoms approach each other.
Then for an entrance channel with no relative orbital angular momentum $\ell$, an $s$-wave channel,
coupling occurs only to closed channels with non-zero partial wave.  As shown in Fig.~\ref{fig:simpleFR}a) 
a non-zero partial wave leads to a centrifugal barrier for the closed channel potentials.

Figure \ref{fig:simpleFR}b) shows the scattering length, $a$, as a function of collision energy $E$ in the presence of a 
Feshbach resonance with resonance energy $E_{\rm res}$. The ($s$-wave) scattering length is a convenient measure of the strength
of the atom-atom interactions at small collision energies. In the limit of zero collision energy the total elastic cross-section $\sigma=4\pi a^2$. Quantum mechanics allows $a$ to be either positive or negative.
Moreover,  the scattering length can depend on collision energy as shown in Fig.~\ref{fig:simpleFR}b) by a distinctive Fano profile for $E\approx E_{\rm res}$.
Away from the resonance the scattering length approaches a background value.

\begin{figure} 
\includegraphics[scale=0.30,trim=0 0 0 0,clip]{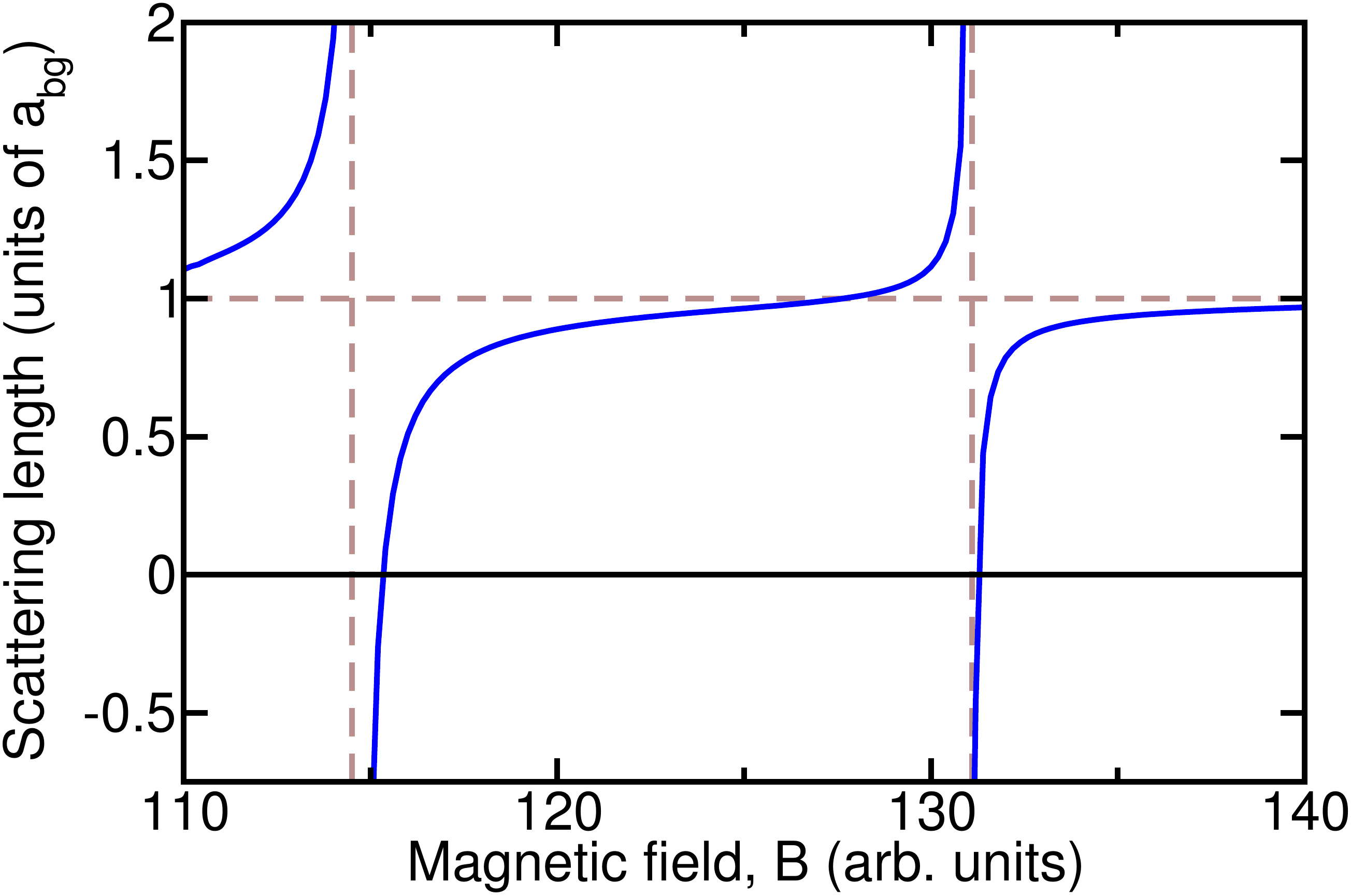}
\caption{Example of a scattering length at zero collision energy as a function of magnetic field. The figure
shows two resonances, one broader than the other, over a small magnetic field region (in arb.~units). The scattering length is scaled in units of the positive background value, a$_{\rm bg}$, away from both resonances. }
\label{fig:avsB}
\end{figure}

Figure~\ref{fig:avsB} gives a second view of the magnetic Feshbach resonance. In this figure the scattering length is shown as a function of magnetic field for zero collision energy.  A resonance occur whenever the scattering length or effective size of the atom goes 
through infinity.
The analytic form for an individual resonance at zero collision energy was derived in \cite{Moerdijk1995} and given by
\[
   a(B)= a_{\rm bg} \left( 1 - \frac{\Delta}{B-B_{\rm res}} \right)\,,
 \]
where $a_{\rm bg}$ is the background scattering length, $B_{\rm res}$ is the magnetic field position of the resonance, and $\Delta$ is the
magnetic width. In fact, it can also be shown that over a sufficiently small magnetic field range $E_{\rm res}(B)=\mu_{\rm res} (B-B_{\rm res})$, where $\mu_{\rm res}$ is the magnetic moment difference of the resonant bound state and the open channel.

\subsection{Role of Feshbach resonances in ultracold collisions}

The crucial role that magnetic Feshbach resonances play  can be understood
from typical densities and temperatures of dilute quantum gases in weak
(harmonic) dipole traps created by focussed laser beams.  The density for
the dilute ultracold gas of atoms varies from $n=10^{12}$ to 10$^{15}$
atoms per cm$^3$, which gives a mean interparticle separation  of around
100 nm. The temperatures are within 1 nK to 1 $\mu$K corresponding
to de Broglie wavelengths $\Lambda_{\rm dB}$ of order of the mean
separation. At phase space densities $n\Lambda_{\rm dB}^3$ of order one
a gas of bosonic atoms forms a Bose-Einstein condensate (BEC). At the
same time the background  $s$-wave scattering length is of order 5 nm,
much smaller than the average separation between the atoms. It can be
tuned via magnetic Feshbach resonances, to much larger values  leading
to strongly interacting quantum gases.

It is also worth noting that typical Zeeman energies, $E_{\rm Zeeman}$,
are of order $k_B\times 100$ $\mu$K  for a magnetic field strength
of order 1 Gauss. Here $k_B$ is the Boltzmann constant.  Only for
much weaker magnetic fields is the Zeeman energy of order of  typical
temperatures.  Magnetic fields of order $10^3$ G are routinely created.
The same analysis shows that for a 1 G change in the magnetic field the
resonance energy $E_{\rm res} (B)$ can change by $k_B\times 100$ $\mu$K.

Feshbach resonances play a much larger role than just being able to
enhance the interaction strength.  They are also used to create a BEC
of  weakly-bound molecules \cite{Kohler2006}. One such bound state is
shown as level $|v'\rangle$ in Fig.~\ref{fig:simpleFR}.  This bound
state can be further stabilized by conversion to a deeply-bound molecule by two- or more-photon
Raman transitions \cite{Science08,Danz2008}.  Finally, three-body Efimov physics
\cite{Kraemer2006} can be explored.

\subsection{Short history }

The first theoretical prediction of Feshbach resonances in collisions
of ultracold cesium atoms was published by Tiesinga {\it et al.}~in
1993 \cite{Tie93}, while the first observation of a Feshbach resonance
in ultracold collisions of sodium atoms was published by Ketterle's
group  in 1997 \cite{Ketterle98}.  The impact of Feshbach resonances in
quantum degenerate alkali-metal gases has been broadly discussed over
the past two decades.  Theoretical concepts of the production of cold
molecules via magnetically-tunable Feshbach resonances  was presented
in Ref.~\cite{Kohler2006}. It was followed by an extensive Review
\cite{Chin2010} that focused on Feshbach resonances as the essential tool
to control the interactions between ultracold alkali-metal atoms. This
included a discussion of a numerous experimental methods to detect
resonances as well as a discussion of their  applications.  Reference
\cite{Chin2010} also provided a description of early history behind the
phenomena of resonant coupling between a bound state and a continuum and
the observation of asymmetric Fano profiles \cite{Fano1961} due to quantum
interference in photo-ionization and absorption spectra.  An excellent
review on recent advances on Bose-condensed quantum gasses of ultracold
dipolar atoms and molecules can be found in Ref.~\cite{Pfau2009}

\subsection{Anisotropic nature of Feshbach resonances}

In this Review we explore the anisotropic nature of Feshbach resonances
in the collision between ultracold highly-magnetic atoms. Highly-magnetic
atoms are atoms with an  electronic ground state, which has a large total
angular momentum $\vec \jmath$ and thus a large magnetic moment $g\mu_B
j$, where $g$ is the atomic $g$ factor and $\mu_B$ the Bohr magneton.
Examples of such atoms are chromium and the lanthanides  erbium and
dysprosium with an angular momentum  $j$=3,  6, and 8, respectively.
In addition, we will describe recent efforts to observe anisotropic
resonances in collision of meta-stable atoms with non-zero electronic
orbital angular momentum $L$.  We will  contrast such resonances to those
observed in alkali-metal atom collisions, where the broadest (strongest)
Feshbach resonances are hyperfine induced and the resonant bound states
do not rotate.

Atoms have a large magnetic moment when several of the electrons in
open electron shells are aligned, either via their spin or their orbital
angular momentum. In chromium the magnetic moment is solely due to the
alignment of the spin of the six electrons in the open 3d$^5$ and 4s
shells. Their total orbital electron wavefunction is spherical (an $S$
state). The most intriguing magnetic atoms  are the submerged-shell
lanthanide atoms. They have an electronic configuration with an open
inner 4f shell shielded by a closed outer
6s$^2$ shell. Their magnetic moment is also due to alignment of electron
orbital angular momenta so that the orbital electron wavefunction becomes
non-spherical.

Interactions between magnetic atoms are orientation
dependent or anisotropic.  At room temperature anisotropic interactions
are much smaller than kinetic energies and other major interactions
between the atoms and, therefore, can be ignored. The situation is different
for an ultracold gas.  Reference \cite{Stuhler2005}, for
example, demonstrated that the anisotropy due to the magnetic  dipole-dipole
interaction between ultracold chromium atoms leads to an anisotropic
deformation of a Bose condensate. Furthermore,
there is a strong evidence that anisotropy plays a dominant role
in collisional relaxation of ultracold atoms with large magnetic moments 
\cite{Krems2004,Weinstein2009,Doyle,Pfau2003,Hancox2004,BLev2010,Kotochigova2011}.

The density of Feshbach resonances as a function of magnetic field is for some highly-magnetic 
atomic species so high that statistical interpretations of the resonance spectrum become necessary. 
Originally, such statistical theories described level distribution in nuclear physics 
\cite{Wigner1951,Dyson1963}, or Rydberg levels in spatial-dependent magnetic fields \cite{Bohigas1984}.

The remaining part of this review is setup as follows. In
Sec.~\ref{sec:long} we start by describing the long-range interatomic
interactions that control the origin of the Feshbach resonances. We focus
in particular on the interplay between the isotropic and anisotropic
interactions.  In Sec. \ref{sec:cr} we describe the role of anisotropic
dipolar interactions on Feshbach resonances in the collisions of
atomic chromium.  We also describe some of the applications of
resonances in the context of many-body physics.  Feshbach resonances
in magnetic lanthanide-atom collisions are non-perturbative in the
anisotropic interactions and will be discussed in Sec.~\ref{sec:dy_er}
for dysprosium and erbium. For erbium resonances
the connection to statistical interpretations of resonance locations
will be established as well.  Resonances in collisions between atoms in
meta-stable states will be discussed in Sec.~\ref{sec:meta}.  We conclude
in Sec.~\ref{sec:conclude}.

\section{Basic physics of atomic interactions}\label{sec:long}

\subsection{Isotropic interactions}

Most current ultracold-atom experiments  use alkali-metal atom gases,
which have only one open valence electron shell. In fact, this shell
is an $s$ orbital containing one electron. The  bond between atoms with
such valence configuration is isotropic. Additionally, for internuclear
separations $R$, where the atomic electron clouds do not significantly
overlap, this bond is characterized by the isotropic van-der-Waals
interaction \[ V(\vec R) \to \frac{C^{\rm iso}_6}{R^6} \quad {\rm
for}\quad R\to\infty\,, \] where $C^{\rm iso}_6$ is the isotropic
van-der-Waals coefficient.  Moreover, it was quickly realized that  as
this interaction energy decays relatively fast with $R$
the complete potential, both short and long range, for many purposes,
such as the modeling of quantum degenerate gases,
can be replaced by a contact delta-function interaction
 \[
 V_{\rm delta}(\vec R) = g_0\,\delta(\vec R)\frac{\partial}{\partial R}R\,,
\] 
with strength $g_0$. This interaction does again not depend on the angle of
approach and is thus isotropic.  Its strength is chosen in such a
way that  both potentials, $V(\vec R)$ and $V_{\rm delta}(\vec R)$,
have the same scattering phase shift $\eta(k)\to-a_sk$ for ultracold
collision energies $E$, where $a_s$ is the $s$-wave scattering length
and wave number $k$ is defined by  $E=\hbar^2k^2/(2\mu_r)$.  This then
lead to $g_0=(2 \pi\hbar^2/\mu_r)\times  a_s$ and $\mu_r$
is the reduced mass of the atom pair.

\subsection{Anisotropic interaction}

In contrast, interactions between atoms with a large permanent magnetic dipole moment
are controlled by anisotropic forces. This anisotropy is present at both short and long range interatomic separations,
but is most easily explained for large separations in terms of the three contributing forces.
They are the magnetic dipole-dipole 
$V_{\mu\mu}(\vec R)$, van-der-Waals dispersion $V_{\rm vdW}(\vec R)$, and  quadrupole-quadrupole 
$V_{QQ}(\vec R)$ interaction potentials, respectively. 
(For details on the short-range potentials see Ref.~\cite{Petrov2012}.)

\begin{figure}
\includegraphics[width=0.25\textwidth,trim=230 180 180 120,clip]{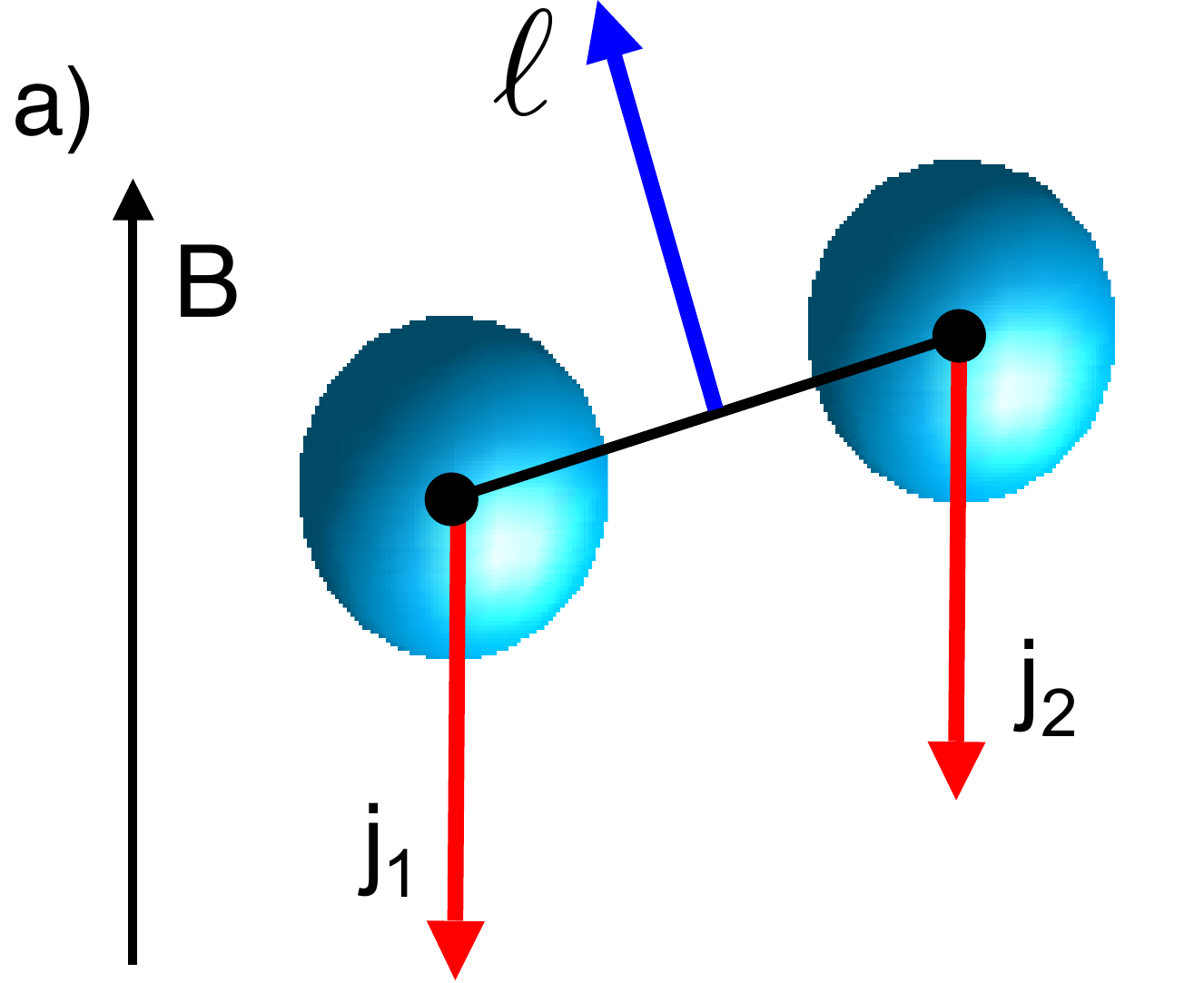}
\includegraphics[width=0.22\textwidth,trim=0 0 0 0,clip]{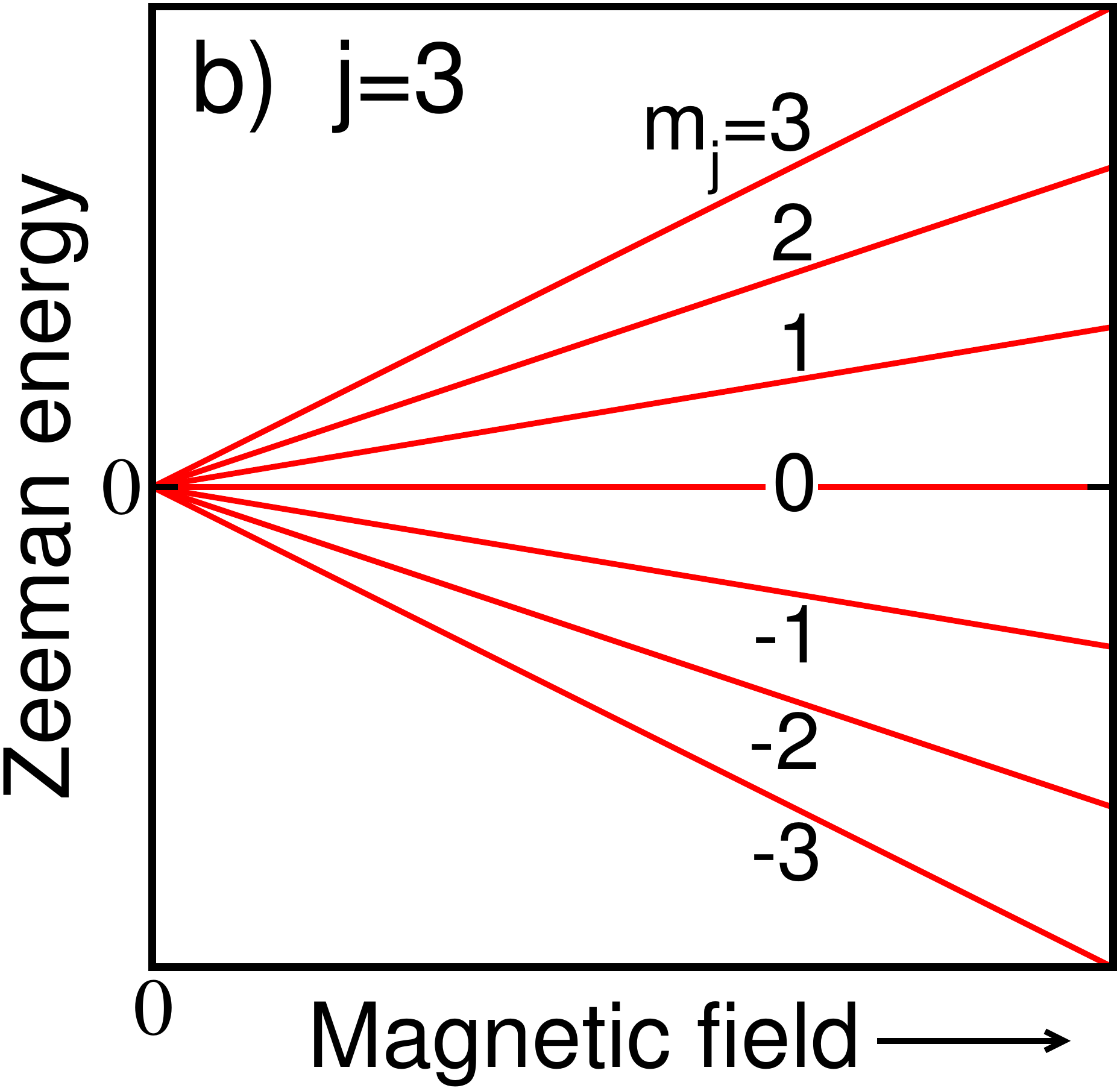}
\caption{a) A schematic of the angular momenta to describe the collision between bosonic magnetic atoms.
b) the Zeeman energy of the magnetic sublevels of a spin-3 magnetic atom as a function of field strength.}
\label{vectors}
\end{figure}

The natural starting point for a collision of bosonic magnetic atoms in a magnetic field are the
orthonormal basis states $|j_1m_1\rangle | j_2 m_2\rangle Y_{\ell m}(\theta,\phi)$, where the kets
$|j_i m_i\rangle$  are the electronic wavefunctions of  atom $i=1,2$ with total atomic angular momentum $\vec\jmath_i$ and projection
$m_i$ along the direction of the magnetic field. The spherical harmonics $Y_{\ell m}(\theta,\phi)$ describe the rotational wavefunction
of the two atoms, where the angles $\theta$ and $\phi$ orient  the internuclear
axis relative to  the magnetic field direction and $\vec \ell$ is the relative orbital angular momentum (also known as the partial wave).
Note that when both bosonic atoms are prepared in the same spin state, only channels with even values of $\ell$ are allowed.
Figure \ref{vectors} shows a schematic picture of these angular momenta as well as the linear-dependence of the Zeeman energy of
the atomic sublevels in a magnetic field.

\begin{figure*} 
\includegraphics[width=0.3\textwidth,trim=60 0 130 0,clip]{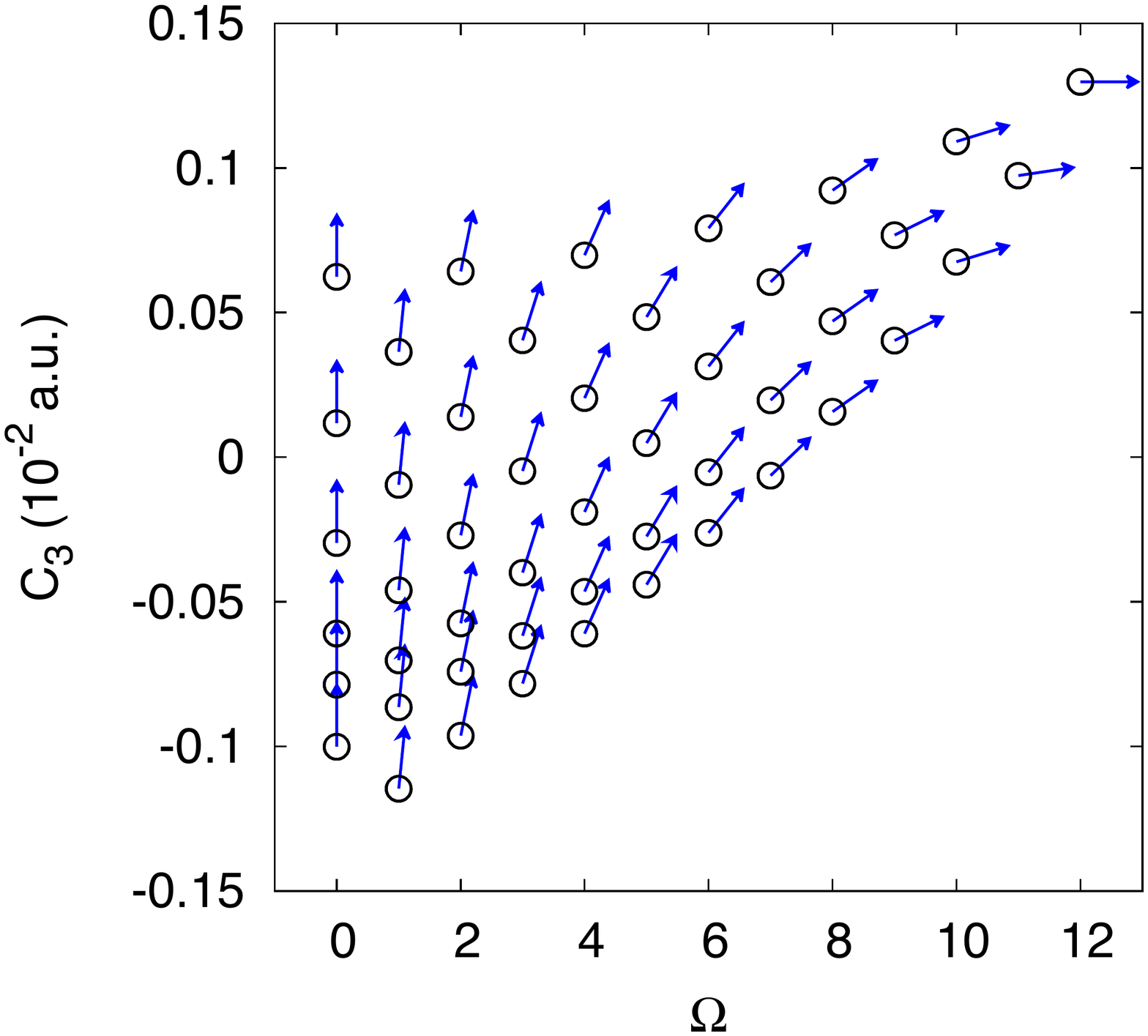}
\includegraphics[width=0.3\textwidth,trim=60 0 130 0,clip]{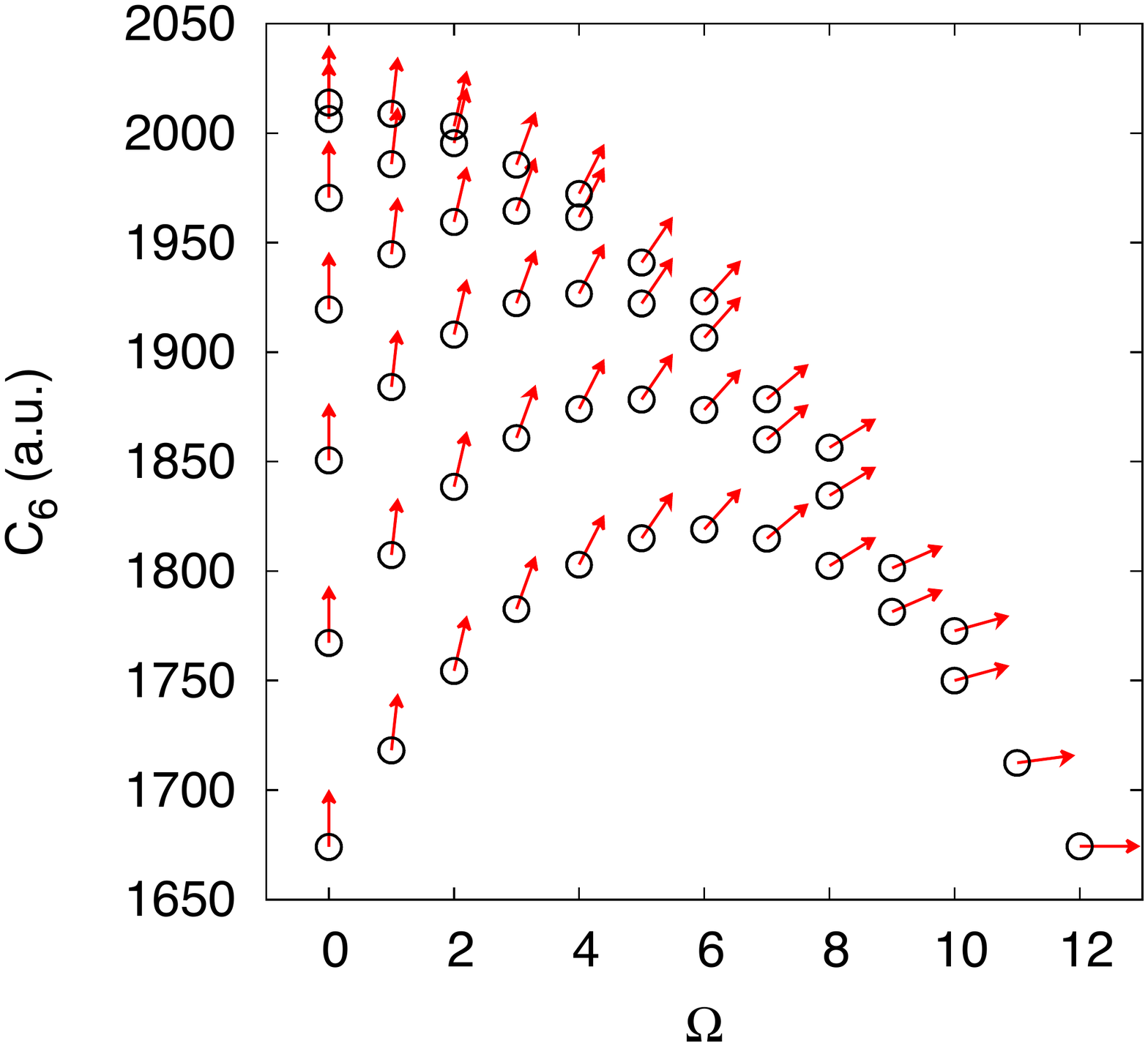}
\includegraphics[width=0.3\textwidth,trim=60 0 130 0,clip]{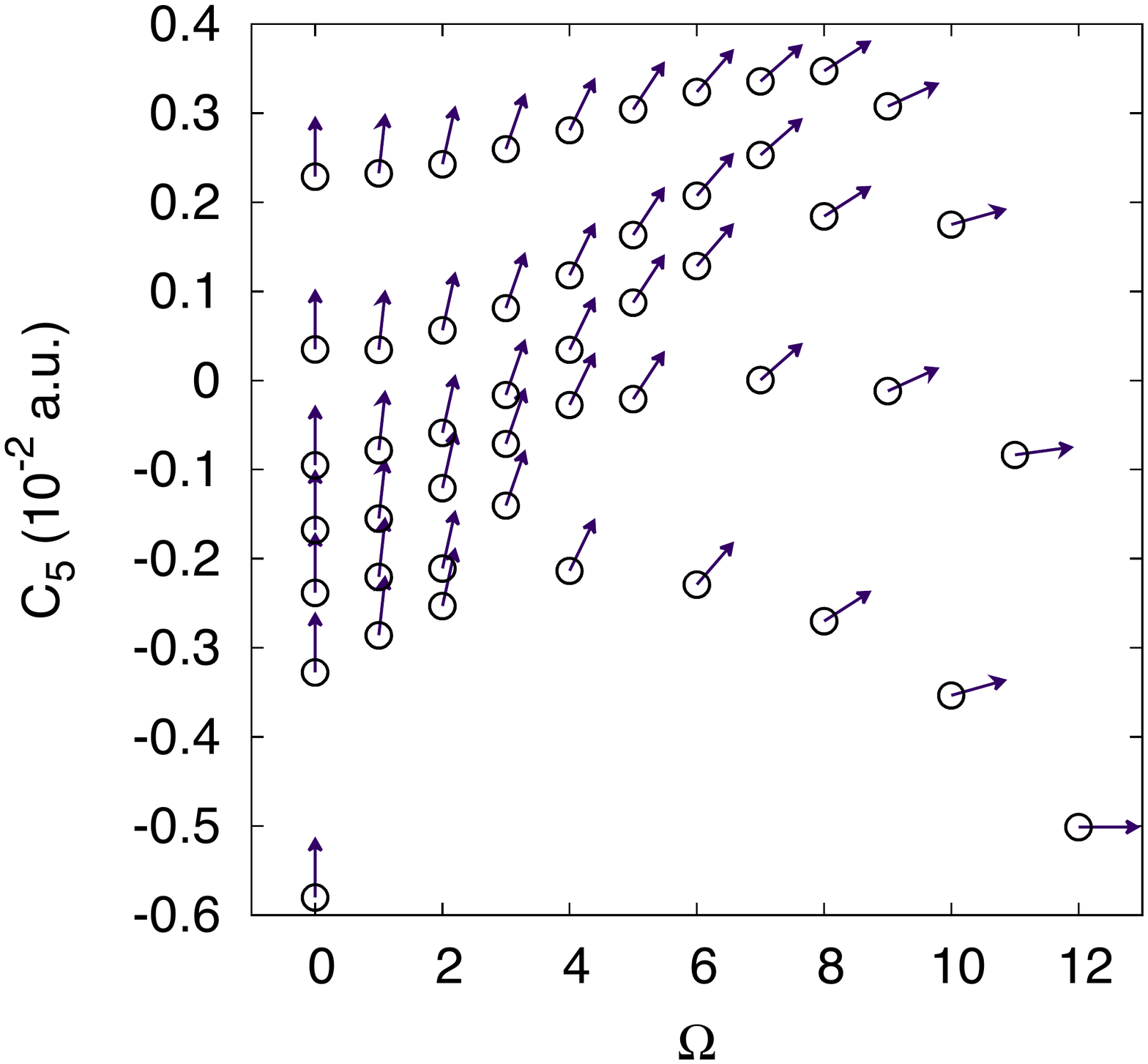} 
\caption{From left to right the {\it gerade} adiabatic $C_3$, $C_6$ and $C_5$ coefficients in atomic units for the interaction between two ground state $^3{\rm H}_{j=6}$ Er atoms 
as a function of the projection $\Omega$ of the total angular momentum $\vec J$ on the interatomic axis. For each $\Omega$ there are approximately $(12-|\Omega|)/2$ {\it gerade} adiabatic coefficients.
Since interactions between atoms are orientation dependent, the arrows on the graph symbolize the anisotropy of the interactions. }
\label{C365}
\end{figure*}

\subsection{Theoretical background for interactions between magnetic atoms}

The three long-range interactions can be systematically represented in terms of tensor operators that describe the coupling between the three 
angular momenta $\vec\jmath_1$, $\vec\jmath_2$, and $\vec\ell$. Following Ref.~\cite{Petrov2012} we have
\begin{eqnarray}
V_{\mu\mu}(\vec R)&=&\frac{c_{\mu\mu}}{R^3} \sum_q (-1)^q
C_{2,-q}(\theta,\phi) T^{(2)}_{2q} ,  \label{long1}
\end{eqnarray} 
\begin{eqnarray}
V_{\rm vdW}(\vec R)&=&-  \!\!\!\! \sum_{k=0,2,4;i}  \frac{c^{(i)}_{k}}{R^6} \, \sum_q (-1)^q 
C_{k,-q}(\theta,\phi)T_{kq}^{(i)} ,  \label{long2} 
\end{eqnarray} 
and
\begin{eqnarray}
V_{QQ}(\vec R) &=& \frac{c_{QQ}}{R^5} \sum_q (-1)^q 
         C_{4,-q}(\theta,\phi) T^{(1)}_{4q}  , \label{long3}
\end{eqnarray} 
where operators $C_{kq}(\theta,\phi)=\sqrt{4\pi/(2k+1)}Y_{kq}(\theta,\phi)$ mix different partial waves $\ell$.
(Examples of this function are $C_{00}(\theta,\phi)=1$ and $C_{20}(\theta,\phi)=
(3\cos^2\theta-1)/2$.)  
The  spherical tensor operators $T_{kq}^{(i)}$ of rank $k$ and component $q$
describe couplings between the atomic angular momenta  $\vec\jmath_1$ and $\vec\jmath_2$
and are given by 
\begin{equation}
T^{(1)}_{00} = I ,\quad
T^{(2)}_{00} =[j_1 \otimes j_2 ]_{00} \,, 
\quad T^{(2)}_{2q} =[j_1 \otimes j_2 ]_{2q} \,,  \label{AMquart} 
\end{equation}
and
\begin{eqnarray}
&&T^{(1)}_{2q} = [j_1 \otimes j_1 ]_{2q} + [j_2 \otimes j_2 ]_{2q} \,,\label{quart}\\                                            
&&T^{(1)}_{4q} = \left[ [j_1 \otimes j_1 ]_2 \otimes  [j_2 \otimes j_2 ]_2 \right]_{4q}  \,,  \nonumber\\  
&&T^{(3)}_{2q} = \left[ [j_1 \otimes j_1 ]_2 \otimes  [j_2 \otimes j_2 ]_2 \right]_{2q}  \,,\nonumber \\ 
&&T^{(3)}_{00} = \left[ [j_1 \otimes j_1 ]_2 \otimes  [j_2 \otimes j_2 ]_2 \right]_{00} \,, \nonumber
\end{eqnarray} 
where $I$ is the identity operator and $ [j_1 \otimes j_2]_{kq}$
denotes a tensor product of angular momentum operators $\vec \jmath_1$
and $\vec \jmath_2$ of atoms 1 and 2 coupled to an operator of rank $k$ and component
$q$ \cite{Kokoouline2003}.  The higher-order tensor operators are constructed in an
analogous manner.  The coefficients $c_{\mu\mu}$, $c_{QQ}$, and $c^{(i)}_k$ are the strengths of the individual terms.

Many of the tensor operators in Eqs.~(\ref{long1})-(\ref{long3})  have a straight-forward interpretation.
Firstly, a contribution is anisotropic when it contains a $C_{kq}(\theta,\phi)$ with non-zero rank $k$.
Moreover, to first-order perturbation theory in the interactions, the
projections $m_1$, $m_2$, and $m$  and partial wave $\ell$  can change up to 2 units
due to the magnetic dipole interaction and up to 4 units due to the
quadrupole-quadrupole interaction and the anisotropic dispersion potential \cite{Kokoouline2003}.

The van-der-Waals dispersion interaction in Eq.~(\ref{long2}) contains multiple contributions.
The largest by far is the isotropic and spin-indendent term proportional to the identity operator.
The term with $T^{(2)}_{00}$ or equivalently  proportional to $\vec\jmath_1\cdot\vec\jmath_2$ induces 
spin-spin coupling without affecting the relative orbital angular momentum.
The dispersion term proportional to $T^{(2)}_{2q}$ in Eq.~(\ref{AMquart}) can be recognized as describing the
same coupling between angular momenta as  the magnetic dipole-dipole interaction.  The van-der Waals contribution,
however, decays as $\propto 1/R^6$.

Equation (\ref{quart}) defines  four more  tensors  $T^{(i)}_{kq}$. Each is connected to a term of the dispersion potential and 
is extremely relevant for the interactions between the Dy and Er lanthanide atoms. 
In fact, Ref.~\cite{Petrov2012} showed that, after the term proportional to the spin-independent $T^{(2)}_{00}$,  the largest dispersion term is the one 
proportional to $T^{(1)}_{2q}$. It corresponds to coupling of the quadrupole moment operator $[j_i \otimes j_i]_{2q}$ of atom $i$ to the rotation of the molecule.
For atomic chromium with its spherical electron wavefunction anisotropic dispersion and quadrupole-quadrupole  interactions are zero leaving only the 
magnetic dipole-dipole interaction as an anisotropic interaction.


This description of collisions and interactions between magnetic atoms in terms of tensor operators should be compared to that of collisions between alkali-metal atoms \cite{Kohler2006}.
Alkali-metal atoms have a non-zero nuclear spin $\vec\imath$ and, in addition to the Zeeman interaction,  an atomic hyperfine coupling between electron and
nuclear spin $\propto (\vec \jmath\cdot \vec\imath)$ must be included. On the other hand,  as remarked upon at the beginning of this section, the bond is isotropic
with a van der Waals potential that is fully given by the simplest tensor $-c^{(1)}_0C_{00}(\theta,\phi)T^{(1)}_{00}/R^6=-c^{(1)}_0/R^6$.
The short-range potentials can be succinctly described by $V_{\rm exch}(R) (\vec\jmath_1\cdot\vec\jmath_2)$,
where $V_{\rm exch}(R)$ is the so-called exchange potential, which is an exponentially decaying function of $R$.
The alkali-metal atoms do have a magnetic moment, but their magnetic dipole-dipole interaction is weak.

\begin{figure}
\includegraphics[width=0.48\textwidth,trim=0 0 0 0,clip]{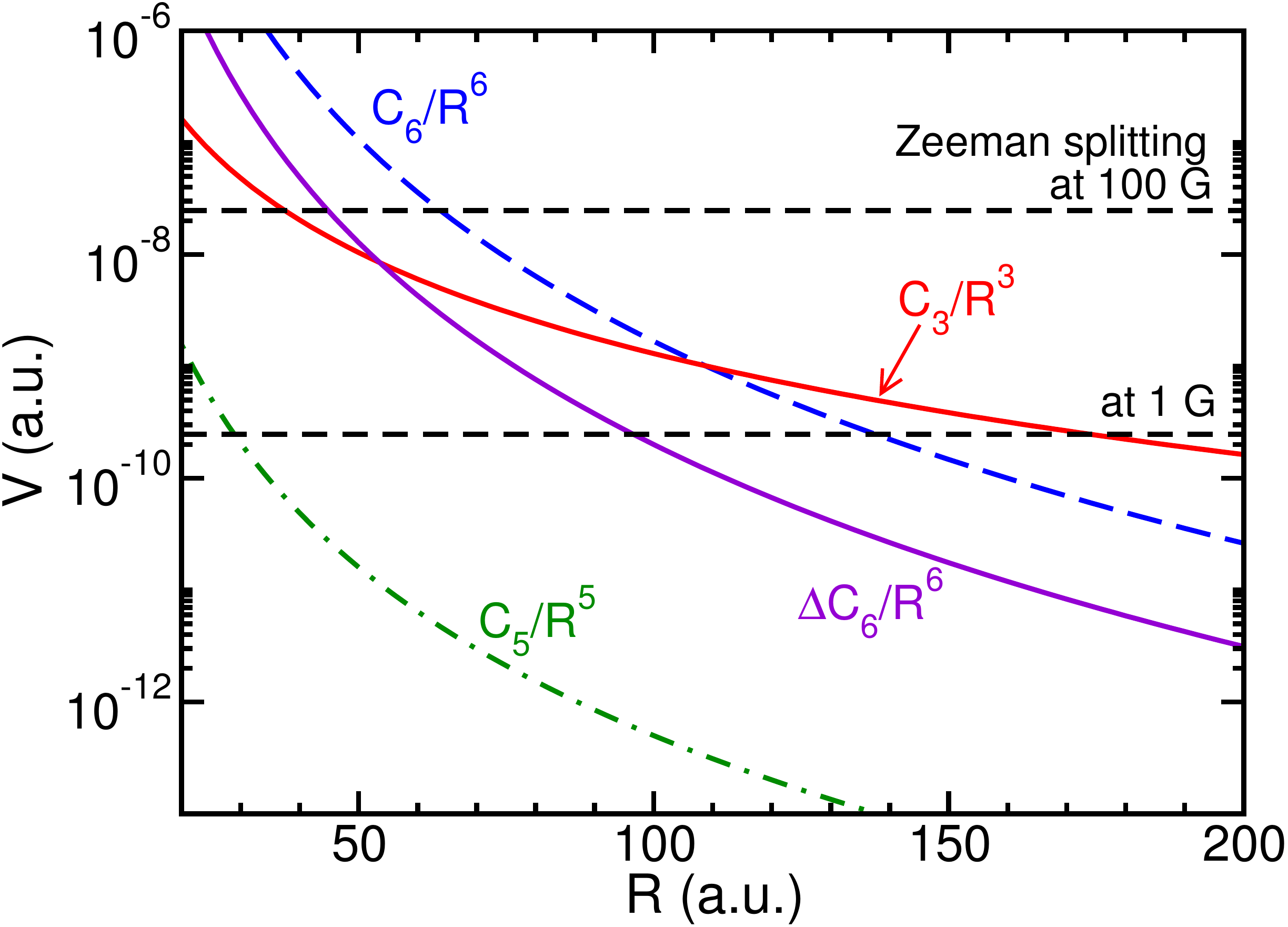}
\caption{Typical long-range interaction potentials and Zeeman level splittings for Er+Er  as a function of interatomic separation in atomic units. 
We have used an isotropic $C_6$=1723 a.u. and a ``mean'' anisotropic  $\Delta C_6$ = 350 a.u. based on Fig.~\ref{C365}.}
\label{ErInt}
\end{figure}

\subsection{Relative size of anisotropic interactions}

An elegant means to represent the character of the anisotropy of the  magnetic dipole-dipole, dispersion, and quadrupole-quadrupole
interactions is to diagonalize  each of Eqs.~\ref{long1}, \ref{long2}, and \ref{long3} in the coupled molecular basis $|(j_1, j_2 )J \Omega\rangle$  
with projection $\Omega$ of  $\vec J$ = $\vec \jmath_1+\vec \jmath_2$ along  the internuclear axis.  This omits couplings between
different projections $\Omega$ due to the rotation of molecule.  Figure~\ref{C365} shows the resulting eigenvalues (multiplied
by $R^n$ with $n=3$, 5, or 6, respectively)  or adiabatic coefficients as a function of $\Omega$ for two Erbium atoms, based on values for 
$c_{\mu\mu}$, $c^{(i)}_{k}$ , and  $c_{QQ}$ from Ref.~\cite{AFrisch2014}. (Authors used experimental  data on atomic transition frequencies and oscillator 
strengths \cite{nist, Lawler2010} and an Er quadrupole  moment  of 0.029 a.u..)

Figure~\ref{C365} shows that the adiabatic coefficients of the three types of interactions have an unique dependence with $\Omega$. 
The values for the dipole-dipole and quadrupole-quadrupole interaction are both positive or negative reflecting the repulsive or attractive 
nature of these interactions depending on the direction at which atoms approach each other. 
The adiabatic van der Waals coefficients $C_6$ are always positive corresponding to predominantly attractive Born-Oppenheimer potentials 
where a larger $C_6$ value implies a deeper  potential. Moreover, they show a smooth nearly parabolic dependence on $\Omega$, 
indicating that one of the rank $k=2$ contributions to the van der Waals potential is the largest anisotropic contribution.
A rank $k=4$ contribution will lead to a quartic dependence with $\Omega$.

To analyze the interplay between different long-range forces in collisions of Er atoms,  Fig.~\ref{ErInt} shows 
the strength of isotropic and various anisotropic potentials as a function of $R$.
In our basis $|j_1m_1\rangle | j_2 m_2\rangle Y_{\ell m}(\theta,\phi)$ the Zeeman
interaction as well as the isotropic dispersion potential (labeled $C_6/R^6$) only shift molecular levels and can not cause inelastic transitions.
The  magnetic dipole-dipole interaction ($C_3/R^3$), anisotropic component of the dispersion potential ($\Delta C_6/R^6$), and  
the negligibly small quadrupole-quadrupole  interaction ($C_5/R^5$)
lead to coupling between Zeeman sublevels. 

For different interatomic separations different interactions dominate. At large $R$ the Zeeman splitting is largest.
When the curves for the magnetic dipole or anisotropic dispersion interaction cross the Zeeman energies $m$-changing
collisions or relaxation can occur.   For small magnetic fields the crossings occur at large
interatomic separations. We also note that for $R > 60 a_0$ the transitions due to the magnetic dipole-dipole interaction
dominate over those of the anisotropic van der Waals interaction.

\section{ Feshbach tuning in collisions of atomic chromium} \label{sec:cr}

\begin{figure} 
\includegraphics[scale=0.35,trim=0 0 0 0,clip]{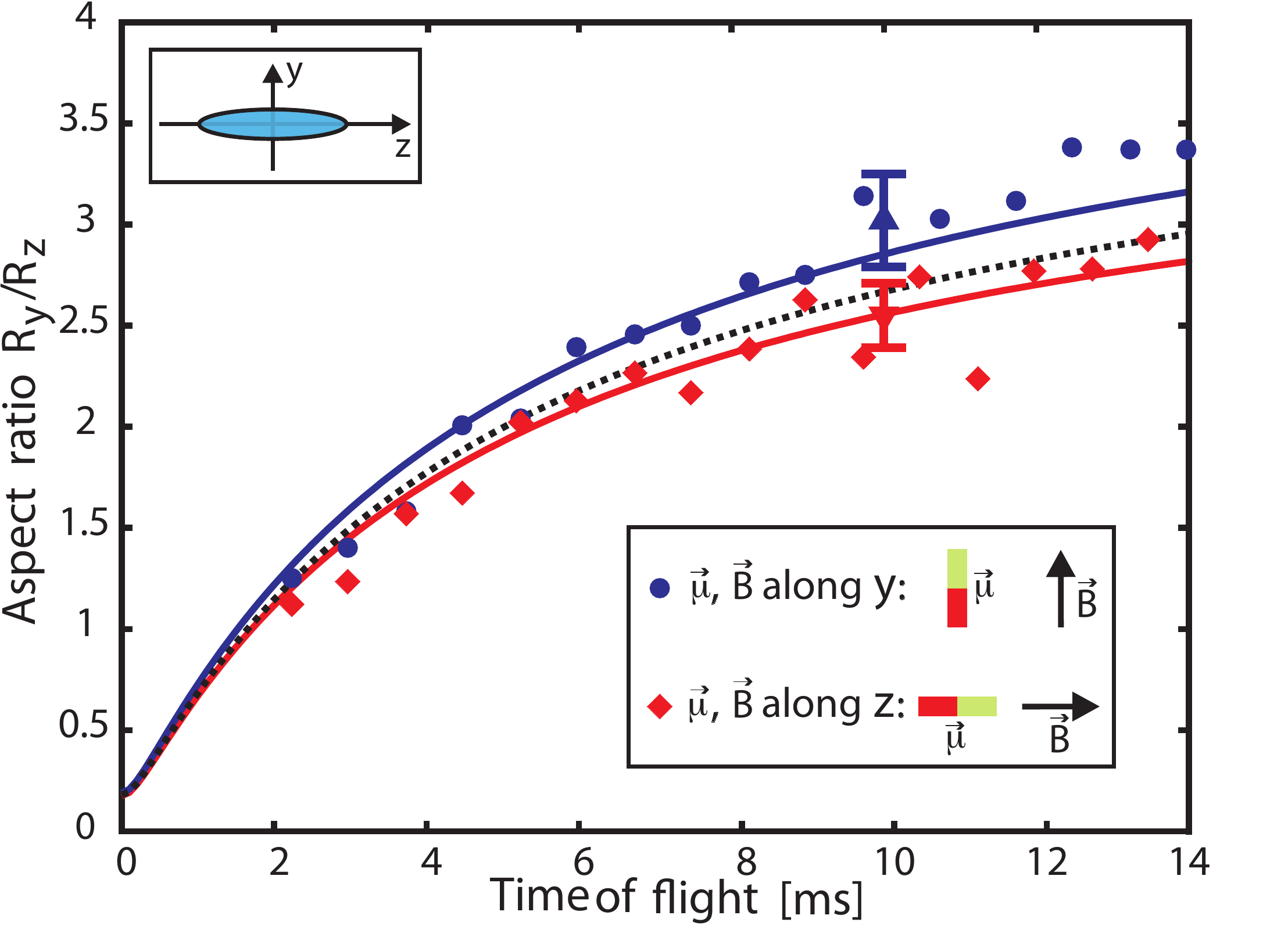}
\caption{
Aspect ratio of a freely expanding Chromium BEC as a function of expansion time for two different directions of magnetization induced by a magnetic field $\vec{B}$ (red and blue curves and markers, respectively). The markers correspond to experimental data. Dashed and solid lines correspond to a theoretical model of the expansion without and with the inclusion of the dipole-dipole interaction, respectively.
Reproduced with permission of Ref.~\cite{Stuhler2005}.}
\label{fig:ddBEC}
\end{figure}

Over the past ten years experimental advances have lead to  better
control of degenerate gases of magnetic $^{52}$Cr atoms in the $^7$S$_3$ ground
state. A large magnetic moment of 6$\mu_B$ initiates a very strong 
anisotropic dipolar interactions that is  36 times stronger than that between alkali-metal atoms.  Developments started at the University of Stuttgart
in the group of T.~Pfau when a Bose-Einstein condensate of Cr atoms
was reported in 2005 \cite{Pfau2005}.  Reference \cite{Stuhler2005} demonstrated, as shown in Fig.~\ref{fig:ddBEC}, that 
by changing the direction of the magnetic field relative to the orientation of a cigar-shaped condensate, dipolar interactions 
modify  the free expansion.

\begin{figure*}
\includegraphics[scale=0.95,trim=0 0 0 0,clip]{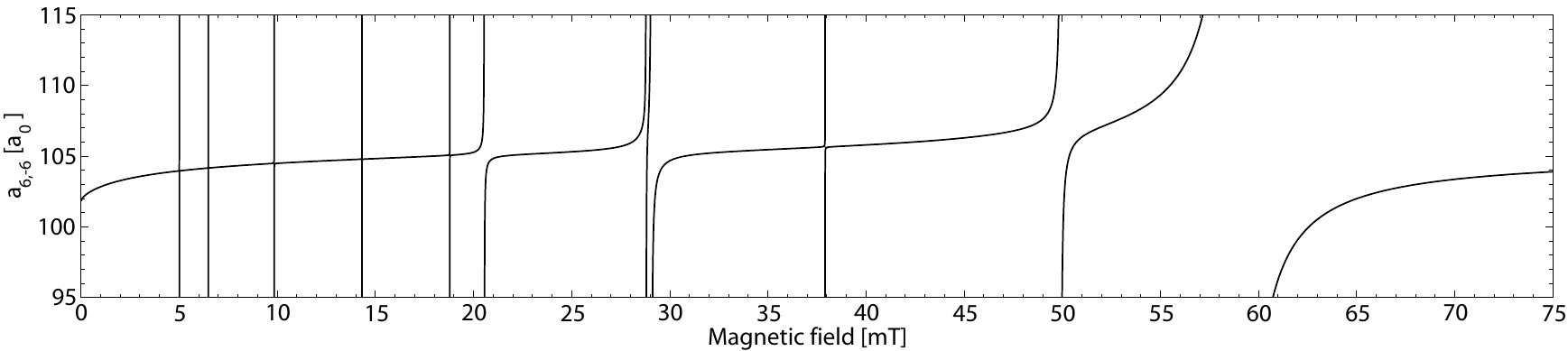}
\caption{
Calculated scattering length of two $m_j = -3$  $^{52}$Cr atoms versus magnetic field where 1 mT = 10 Gauss. Model parameters  
are given in Ref.~\cite{Tiesinga2005}. See Table 1 of this reference for the experimental data. 
Reproduced with permission of Ref.~\cite{Tiesinga2005}.}
\label{fig:FR}
\end{figure*}

The first observation of Feshbach resonances in the collisions of ultracold bosonic  
$^{52}$Cr atoms in the $m_j=-3$ state was reported in Ref.~\cite{Tiesinga2005}. They found 
ten resonances, shown in Fig.~\ref{fig:FR},  between $B$= 4 G and 600 G, leading to
an average density of resonances of $\approx0.02$  per Gauss. 
The zero nuclear spin of $^{52}$Cr and, thus, the absence of a hyperfine Fermi-contact
interaction allowed for an accurate model and identification of the observed resonances using multichannel
scattering calculations.  In fact, the average discrepancy between theoretical and experimental 
resonance positions is only 0.6 G. The dipole-dipole interaction was included in the theoretical modeling
of Ref.~\cite{Tiesinga2005} allowing an accurate discription of the widths of the observed resonances. 
A similar theoretical analysis of the Feshbach resonances observed and characterized in \cite{Tiesinga2005}
was performed in Ref.~\cite{Cote2005}. 

\subsection{Direct evidence for dipolar effects}

\begin{figure}
\begin{center}
\includegraphics[scale=1,trim=150 340 30 200,clip]{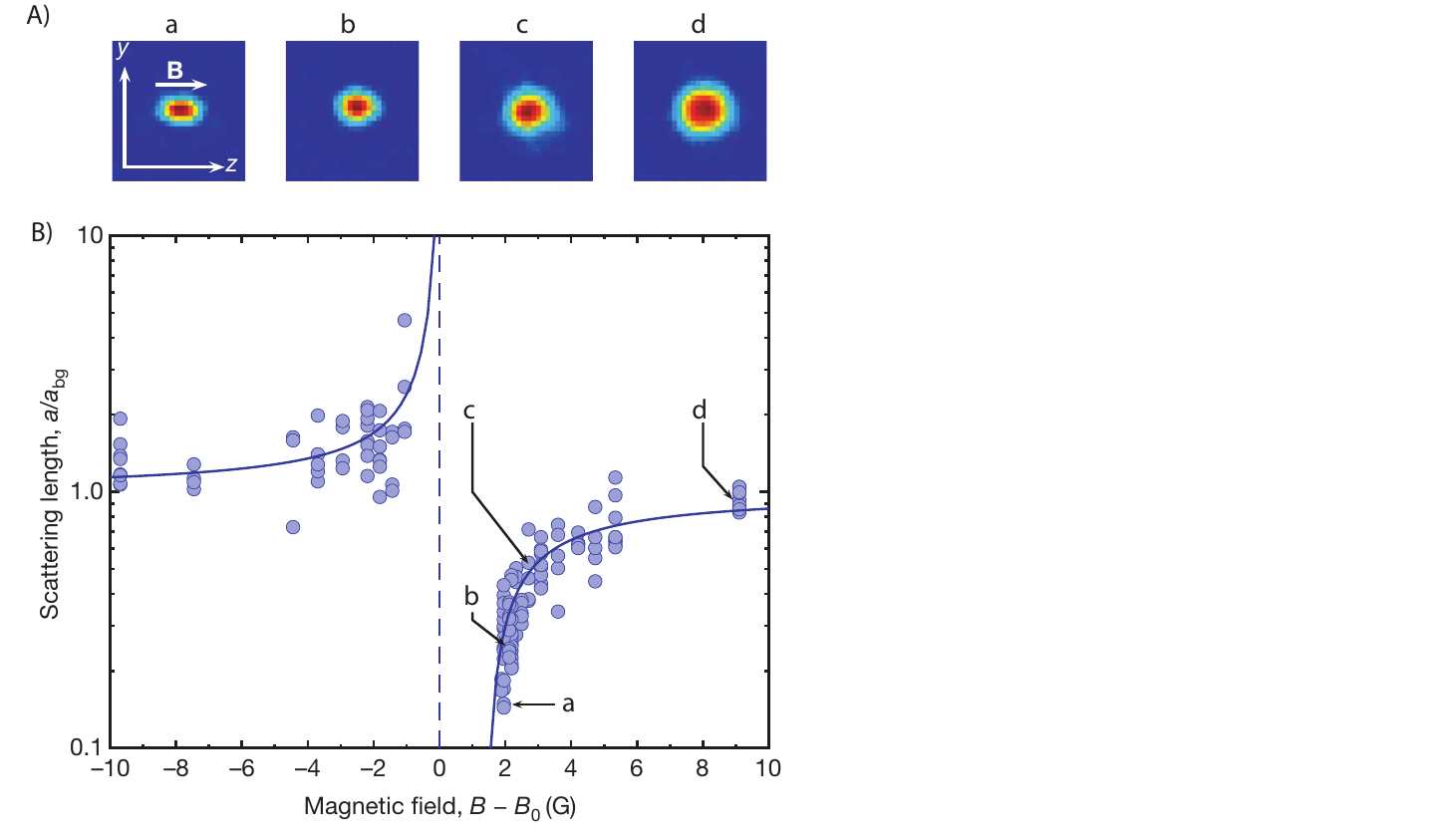}
\end{center}
\caption{
a) Absorption images of a $^{52}$Cr Bose condensate after expansion for four values of the magnetic field. 
b) Measured scattering length of two $m_j = -3$  $^{52}$Cr atoms versus magnetic field 
with $B_0=589$ G. The fields labeled a,b,c, and d correspond to the four magnetic fields in panel a), respectively.
One clearly observes an decrease in size and increase of ellipticity when the scattering length decreases.
Reproduced with permission of Ref.~\cite{Pfau2007}.}
\label{fig:ScatLength}
\end{figure}

Later it was shown that  effects of dipolar  forces in a quantum gas of Cr
can be brought out using Feshbach resonances \cite{Stuhler2010}. The 
broadest resonances in Fig.~\ref{fig:FR}, at $B=589$ G, was
selected for Feshbach tuning of  the $s$-wave scattering length. With
the scattering length set  close to zero, direct evidence for dipolar
effects on BEC was observed \cite{Pfau2007,Koch2008}.  It was shown that the
magnetic dipole-dipole interaction energy can be comparable to the
so-called mean-field energy. Figure~\ref{fig:ScatLength} shows their observation of the scattering
length as a function of the magnetic field near the 589 G Feshbach resonance.  
A characteristic anisotropic $d$-wave
collapse and subsequent explosion, presented in Ref.~\cite{Lahaye2008},
gave further evidence of the relevance of dipole-dipole forces.

Dipolar interactions have other consequences as well.  A chromium
sample prepared in the low-field seeking $m_j=+3$ state  will undergo
strong dipolar relaxation, where, during a collision,  one or both
atoms change its state to a sublevel with a smaller $m_j$. The
decrease in internal energy leads to an increased relative kinetic
energy and the atoms are, typically, lost from the shallow traps in
which the atoms are held.  This relaxation was observed as early as
2003 \cite{Pfau2003}, It was shown that
the cross-section for relaxation scales as the cube of the magnetic
dipole moment.  Additional dipolar relaxation rates were measured in
Refs.~\cite{BLaburthe2010PRA,BLaburthe2011,BLaburthe2011_2}. Furthermore,
they showed that relaxation can be controlled by static and  oscillatory
magnetic fields.

Chromium atoms have also been loaded into optical lattices, periodic potential created counter-propagating laser beams. 
Reference~\cite{Pfau2011} showed that for an one-dimensonal optical lattice the stability of a pan-cake-shaped dipolar 
$^{52}$Cr condensate near $B=589$ G dramatically depends on the depth of the lattice.
The stability measurements were performed at a magnetic field near a Feshbach resonance, where the dipole-dipole 
interaction dominates the short-range isotropic interactions.
Another effect of a strong dipole-dipole interaction is that a gas can be stable in an optical lattice,
but is not during a time-of-flight expansion after the lattice trap is turned off \cite{Billy2012}.
Non-equilibrium quantum magnetism at very small magnetic field strengths was studied in experiments by B.~Laburthe  of
University Paris 13$^{\rm th}$ \cite{BLaburthe2013}.  They showed that non-equilibrium spinor dynamics is modified by the 
non-local inter-site dipole-dipole interactions.

\subsection{Cooling effect of dipolar relaxation}

\begin{figure} 
\begin{center}
\includegraphics[width=0.48\textwidth,trim=100 570 200 100,clip]{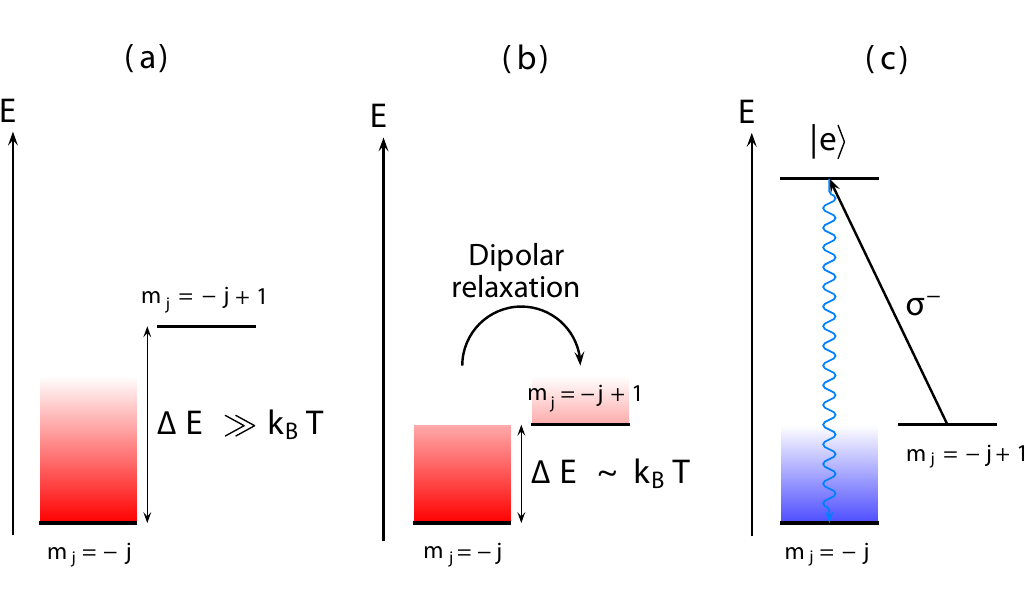}
\end{center}
\caption{Principle of demagnetization cooling.   a) In a 
magnetic field where the Zeeman splitting $\Delta E$ between adjacent Zeeman 
sublevels $m_j$ is larger than the thermal energy $k_B T$,  a gas of magnetic atoms in the 
energetically-lowest $m_j = -j$ state is stable.  There are no inelastic dipolar relaxation processes.
b) By slowly reducing the field strength so that $\Delta E \approx k_B T$, the dipole-dipole induced transitions to 
state $m_j = -j+1$ become allowed and kinetic energy is converted into Zeeman energy. c) By applying an optical pumping pulse 
of $\sigma^-$ polarized light, the cloud can again be polarized but now is at a reduced  temperature. 
The excess Zeeman energy is taken away by the spontaneously emitted photons.
The lowest achievable temperature is of order the photon recoil energy, the energy added by the
optical pumping process. Reproduced with permission of Ref.~\cite{Pfau2009}.
} \label{demag}
\end{figure}

More recently, dipolar relaxation was used to cool
a sample by adiabatic demagnetization. This cooling scheme
was suggested in Ref.~\cite{Pfau_EuroLett} and demonstrated
experimentally in Ref.~\cite{Pfau2006}.  Figure~\ref{demag} shows how inelastic collisions
can be used to implement  adiabatic demagnetization. 
The scheme relies on collisional relaxation in extremely small
magnetic fields, where Zeeman splittings are of order the temperature of the gas,
and on  spin selection rules that can only by achieved by anisotropic interactions, 
where the atomic angular momentum couples to the rotational state of the colliding atoms. 
Recent  research \cite{Volchkov2013} in  demagnetization cooling of  Cr atoms
showed a significant improvement in efficiency over a large temperature range and for high atomic
densities.  The authors discuss  the possibility of achieving Bose-Einstein condensation by demagnetization
cooling of  Dy atoms.   

The role of the dipole-dipole interaction in spinor dynamics, the time evolution
coherences between and populations of $m_j$ sublevels, in a chromium BEC was explored in
Refs.~\cite{Santos2006,Pfau2007PRA}.  The dynamics resembles the Einstein-de Haas effect.
Anisotropic coupling transfers atoms from sublevel $m_j$ to $m_j+1$ leading to the generation of dynamical rotation.
Reference \cite{Kawaguchi2006} showed that the  Einstein-de Haas effect is easily
destroyed  due to the role of Larmor precession in  an external magnetic field.

In summary, the most important feature of a Cr quantum gas is a strong
anisotropic dipole-dipole interaction, based on the large magnetic moment
(electronic spin) of  the Cr atom.  These interactions lie at the heart
of many fascinating effects observed or predicted for a Cr BEC.

\section{Degenerate gases of dysprosium and erbium atoms}\label{sec:dy_er}

\subsection{Collisional properties of submerged-shell atoms}

Over the past decade significant attention has been devoted
to the characterization of the interactions between 
submerged-shell 3d-transition-metal and 4f-rare-earth atoms
\cite{Pfau2003,Hancox2004,Hancox2005,Krems2005,Stuhler2005,Harris2007,Pfau2007,Doyle}.
These atoms have an electronic configuration with an electron vacancy
in the inner shell shielded by a closed outer shell.  It was long
assumed that inelastic, energy-releasing collisions of submerged
shell atoms are substantially suppressed due to shielding caused by
the closed outer-shell electrons.  This effect was first predicted and
demonstrated for collisions between submerged-shell atoms with helium
\cite{Hancox2004,Hancox2005,Krems2005,Tscherbul2009}.  The suppression of inelastic loss
with a cold gas of He atoms allowed for sympathetic cooling of submerged
shell atoms to millikelvin temperatures.  Theoretical analyses by
Ref.~\cite{Buchachenko2005} of experimental measurements \cite{Hancox2004}
explain this low rate by the fact that anisotropy in interactions between
open-shell lanthanide atoms and helium is extremely small.

Measurements \cite{Harris2007,Weinstein2009,BLev2010,Doyle} 
of the spin relaxation rates in collisions of two submerged-shell
atoms, however, have shown no such suppression and, in fact, the rate
coefficients are of the same order of magnitude ($10^{-10}$ cm$^3$/s) as for non-submerged
shell atoms. This implied the presence of additional spin relaxation mechanisms.

Submerged-shell atoms, such as dysprosium Dy($^6$I$_8$) and erbium
Er($^3$H$_6$), focussed on in this Review, do not only have a large
magnetic moment but also large non-zero orbital angular momentum $L$. The
electronic structure of these non-S state atoms  leads to an additional
source of anisotropy in their interactions and it is, for example,
of relevance to determine its effect on Feshbach resonance tuning and
control.

The importance of anisotropy in the interactions of cold atoms with
non-zero angular momenta was first theoretically investigated by
Refs.~\cite{Kokoouline2003,Santra2003,Krems2004,Groenenboom2007}. They have shown
strong evidence  of electronic interaction anisotropy as a leading spin
relaxation mechanism in collisions.

\subsection{Universality in collisions of Dy atoms}\label{sec:dy}

\begin{figure}
\includegraphics[width=0.47\textwidth,trim=0 0 0 0,clip]{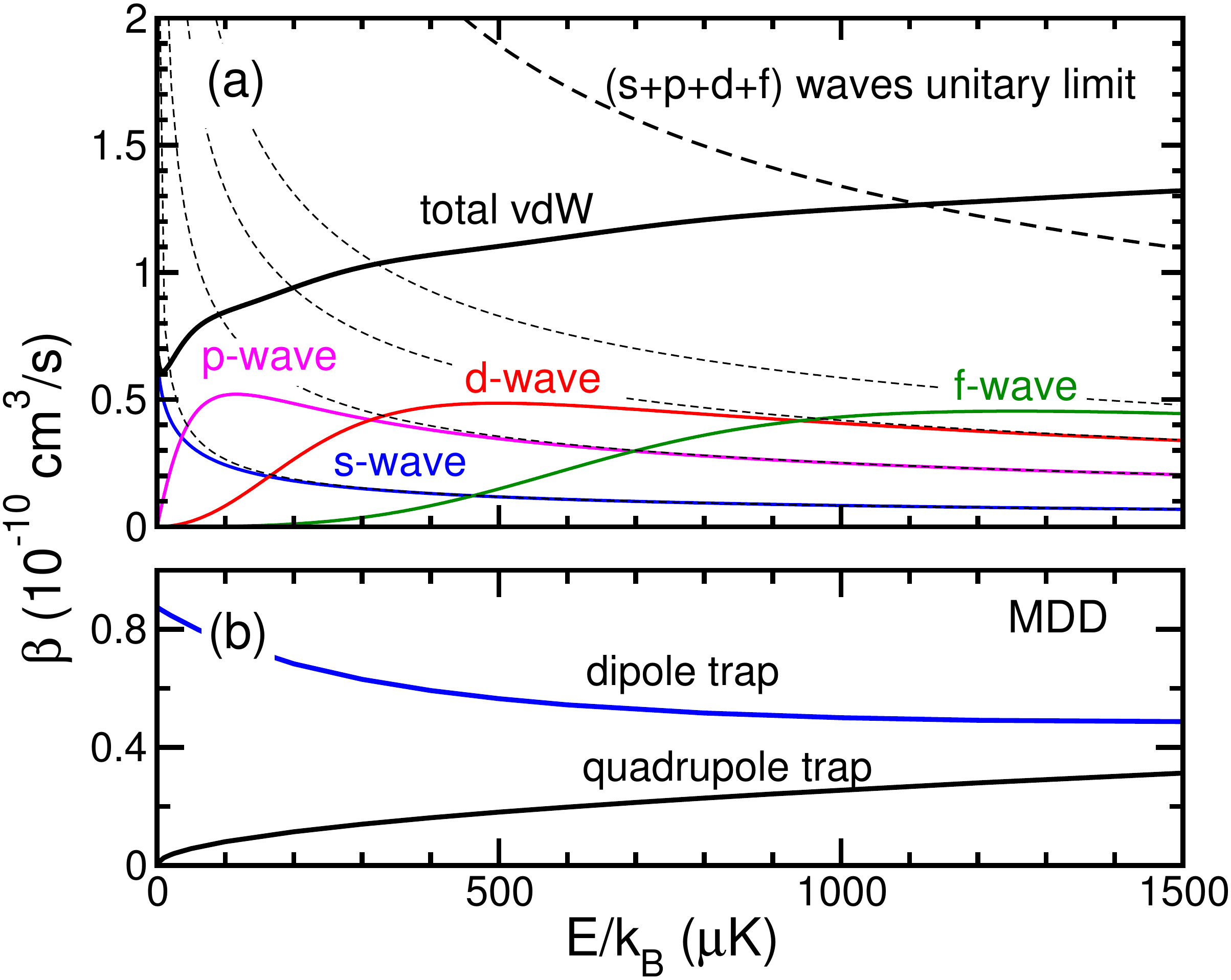}
\caption{The inelastic loss rate coefficient for a non-spin-polarized
sample of ground state $^{164}$Dy atoms as a function of collision
energy based on an universal scattering model for losses due to the anisotropy of the
dispersion potential (panel a) and a Born approximation for losses from the magnetic 
dipole-dipole interaction (panel b).  For the universal model rate coefficients 
for the lowest four partial waves and the summed rate are shown.  The unitary 
limited rate coefficients for each of the four partial waves are plotted as 
dashed lines. The loss rate coefficient for the magnetic dipole-dipole interaction 
is given for a quadrupole as well as a dipole trap with a constant magnetic field of $B$ = 10 G. 
Reproduced with permission of Ref.~\cite{Kotochigova2011}
}
\label{InLoss}
\end{figure}

A wealth of fascinating properties of interacting Dy atoms in
their ground state was revealed by the theoretical analyses of
Ref.~\cite{Kotochigova2011}.  Using experimental data on atomic transition
frequencies and oscillator strength authors constructed 153 interaction
potentials  that dissociate to two ground-state atoms.  Splittings between
these potentials provides an estimate for the strength of the anisotropic
forces that play a crucial role in the alignment of the open 4f-shell
electrons and in sublevel- or $m$-changing relaxation mechanisms.
In addition, the authors used an universal scattering model to study
inelastic scattering and estimate loss rates.  This model, originally
developed in Refs.~\cite{Mies84,Julienne}, assumes scattering from
a single potential $-C_6/R^6 + \hbar^2\ell(\ell+1)/(2\mu_r
R^2)$  for $R>R_c$ and where $C_6$ is equal to the isotropic van der
Waals coefficient.  Atom pairs that reach the critical interatomic
separation $R_c$ undergo $m$-changing collisions with unit probability
independent of scattering energy and partial wave $\ell$.  The model
assumes that for $R>R_c$ coupling due to the anisotropic 
dispersion potential and the dipole-dipole interaction can be neglected.
Figure~\ref{InLoss}a shows inelastic rate coefficients for a gas of
Dy atoms with equal populations in all $m$ sublevels as a function of collisional energy within
this universal model.

Another important anisotropic interaction between Dy atoms that can cause
inelastic losses comes from the magnetic dipole-dipole interactions. The
rate of these losses was estimated in Ref.~\cite{Kotochigova2011} by
using perturbative Born approximation and shown as a function of collision
energy in Fig.~\ref{InLoss}b. These rates were compared to experimental
loss rate measured at a temperature of $\approx$ 500 $\mu$K  confined in
a quadrupole magnetic trap (a trap with zero magnetic field in the center)
in Ref.~\cite{BLev2010}.  Both experiment and theory show rates of the
order of 10$^{-10}$ cm$^3$/s as predicted for other submerged-shell
atoms in Ref.~\cite{Doyle}.

\subsection{Quantum degenerate gas of Dy atoms}

A direct and efficient transfer of atoms into an optical dipole trap allowed
researchers from the University of Illinois and Stanford University to
form a Bose condensate of the bosonic $^{164}$Dy atoms at temperatures below 30 nK
\cite{Lev2011}.  They also cooled fermionic $^{161}$Dy in the presence
of bosonic isotopes to form a Fermi sea of atoms thus realizing a novel,
nearly quantum degenerate dipolar Bose-Fermi mixture \cite{Lev2012}.

\begin{figure}
\includegraphics[scale=0.32,angle=0,trim=0 0 0 0,clip]{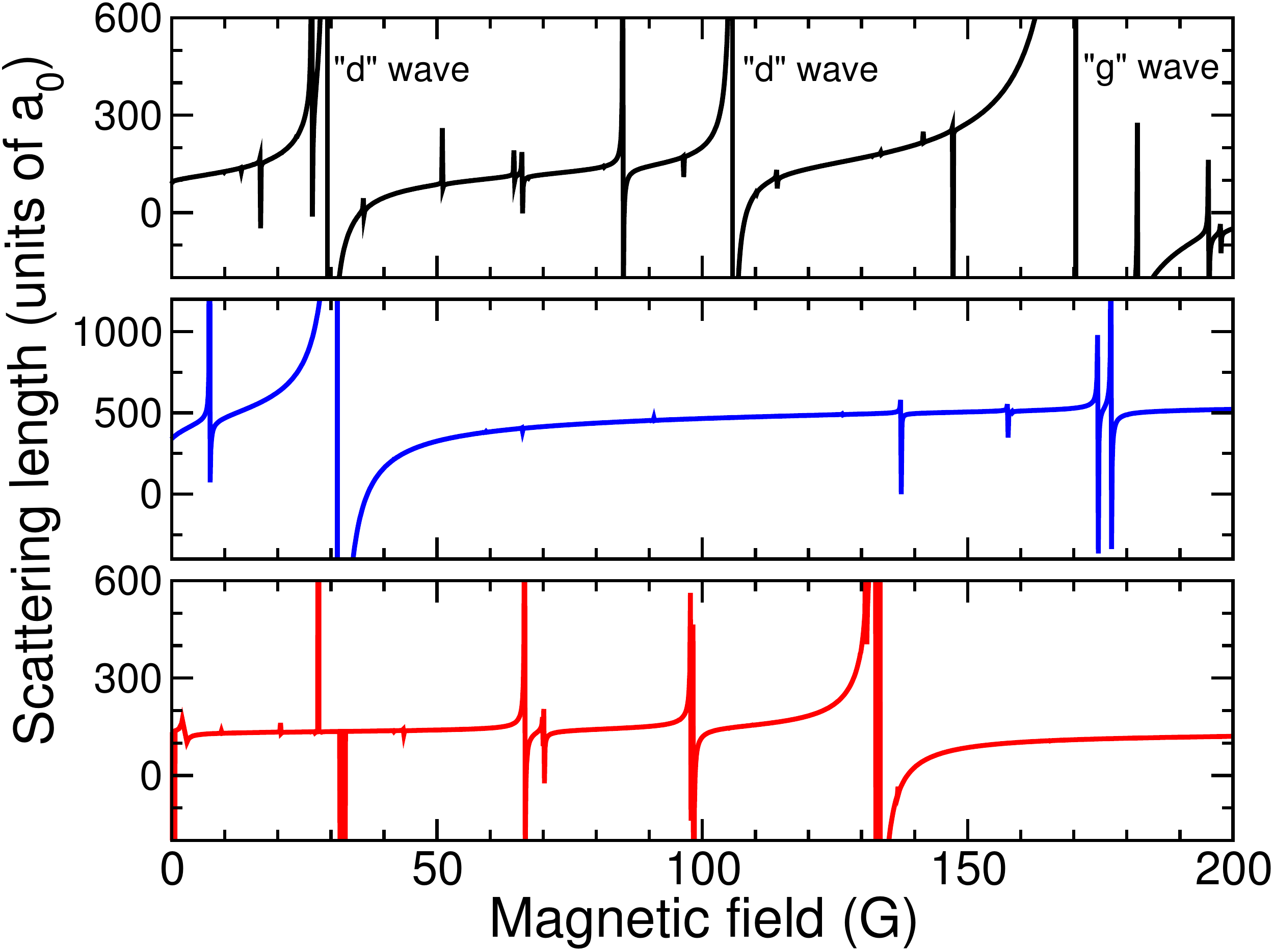}
\caption{Predicted scattering length of $m=-8$ $^{164}$Dy atoms as a
function of magnetic field with and without the magnetic dipole-dipole
or the anisotropic contribution of the dispersion interaction. 
Channels with even partial waves $\ell$ up to 10 are included and a collision energy $E/k_B=30$ nK was used.
The top panel shows the case when all interactions are included.  
For the three broad resonances the first
partial wave for which the resonance appears is shown. The middle and bottom panels are
obtained when the dispersion and magnetic dipole-dipole anisotropy is set
to zero, respectively. Reproduced with permission of Ref.~\cite{Petrov2012}.
}
\label{fig:DyFR}
\end{figure}

Recent theoretical work \cite{Petrov2012} performed a fully
quantum-mechanical scattering calculation of the scattering length and
elastic rates between two ultracold dysprosium atoms in the lowest
Zeeman sublevels $m_j = -8$ and under experimental conditions of
Ref.~\cite{Lev2012}. This investigation predicted for the first time the
existence strong and broad Feshbach resonances in interaction between
bosonic $^{160}$Dy,  $^{162}$Dy, and $^{164}$Dy atoms, which have zero
nuclear spin, for a magnetic field range from zero to 200 Gauss. These
resonances are solely induced by the anisotropy in the long-range
interaction potentials. Without the magnetic dipole-dipole and anisotropic
dispersion potentials in the Hamiltonian resonances do not occur. Both
anisotropies contribute to the appearance of a resonance structure.
Figure~\ref{fig:DyFR} provides an evidence of the direct effect of both
anisotropies on the magnetic-field location of Feshbach resonances.
Switching on and off different parts of Hamiltonian, the researchers observed
a significant change in the resonance distribution.

\begin{figure} 
\includegraphics[width=0.48\textwidth,trim=0 0 0 0,clip]{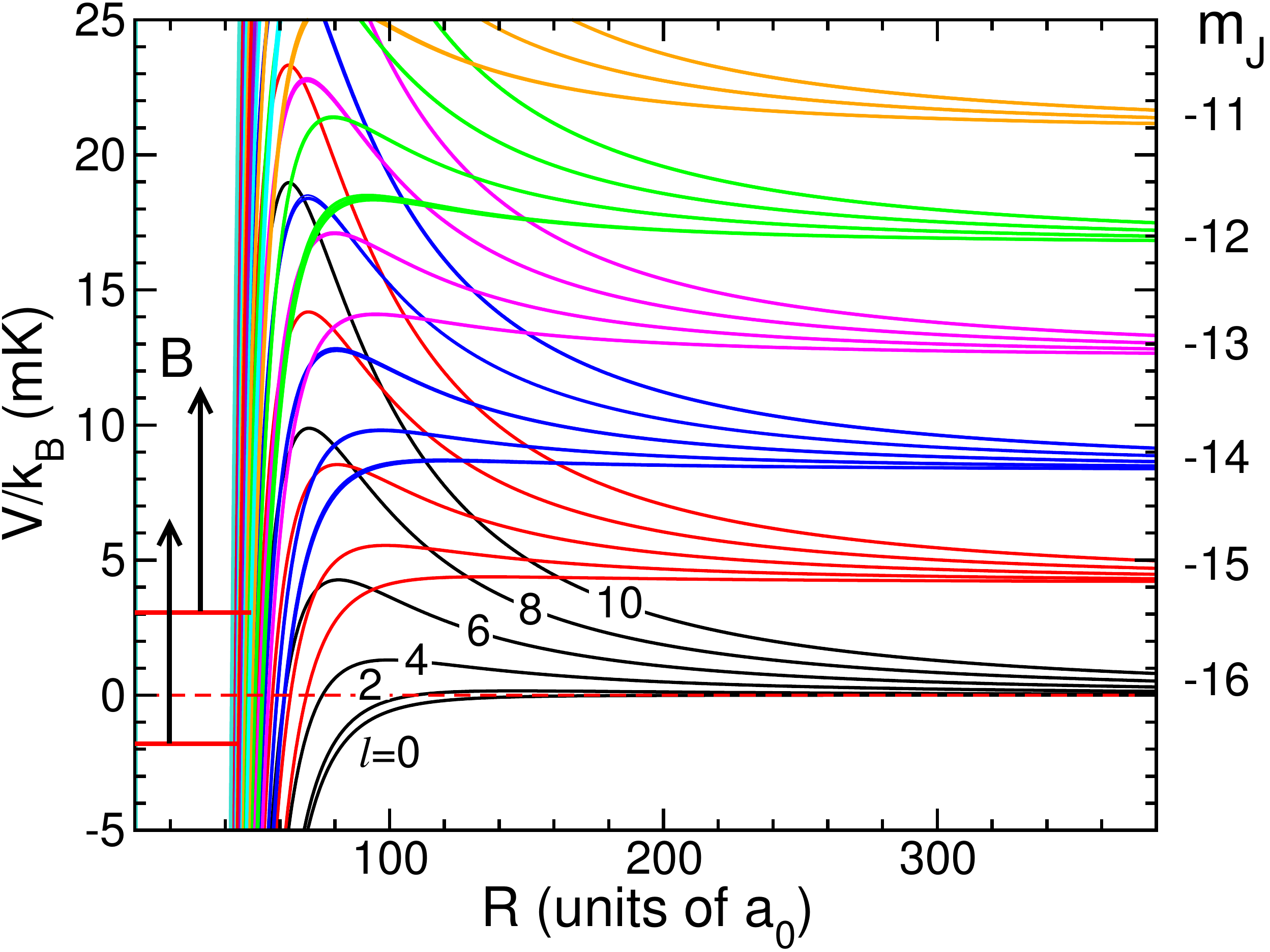}
\caption{Potential energy curves for a
$^{164}$Dy+$^{164}$Dy collision for $B$ = 50 G as a
function of interatomic separation $R$.  The (red) dashed line with zero
energy indicates the energy of the entrance channel.  
The curves are colored by their $m_j=m_1+m_2$ value, while for $m_j=-16$ curves their $\ell$ value is indicated.
Reproduced with permission from authors of Ref.~\cite{Petrov2012}.}
\label{fig:diabatic}
\end{figure}

Despite of the fact that the broadest resonances in Fig.~\ref{InLoss} were
identified as ``d''- and ``g''-wave resonance channels with $\ell$ up up
10 needed to be included to converge the close-coupling calculation. The
long-range potential energy curves of interacting $^{164}$Dy atoms for
a  magnetic field $B = 50$ G are shown as a function of interatomic
separation in Fig.~\ref{fig:diabatic}. For  $R > 200a_0$  the Zeeman
forces dominate the collision dynamics, whereas for $R < 200a_0$
the potential curves of  higher partial waves overlap indicating the
possibility of coupling between potentials.

\subsection{First observation of low-magnetic field Feshbach resonances in Dy collisions} 

\begin{figure}
\includegraphics[scale=0.9,angle=0]{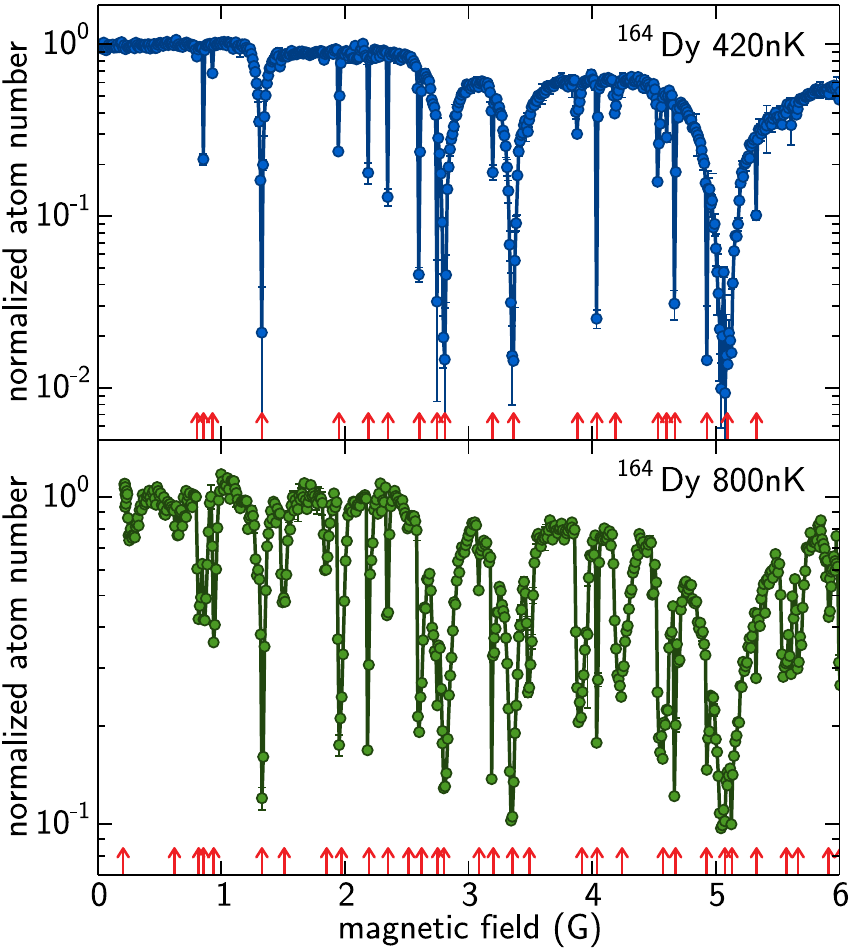}
\caption{Experimental observation of Feshbach resonances in collisions of $m_J=-8$ $^{164}$Dy atoms of Ref.~\cite{BLev2014}.
The top panel shows spectrum obtained at a temperature of $ \approx$ 420 nK. The lower panel displays a spectrum at a higher
temperatures of $\approx$ 800 nK. The arrows indicate the positions of the resonances.
}
\label{fig:FR_Dy}
\end{figure}

Reference~\cite{BLev2014} has recently reported the observation of collisional resonances in trap loss
of spin-polarized Dy atoms in their energetically-lowest Zeeman sublevel. The measurements 
were performed for three bosonic isotops, $^{160}$Dy, $^{162}$Dy, and $^{164}$Dy, and a single fermionic
isotope of $^{161}$Dy. The bosonic atoms were transferred into the $m_J = -8$ magnetic sublevel , while
the fermionic atoms were placed into the lowest hyperfine sublevel $F = 21/2,\,m_F = -21/2$, where  $\vec F$ is the sum of the total electronic 
angular momentum and the nuclear spin. Observation of Feshbach resonances  as shown in Fig.~\ref{fig:FR_Dy} was recorded in the magnetic
field range from 0 G to 6 G, where the resonant density exceeded 3 resonances for Gauss for bosonic  $^{164}$Dy
at a temperature of 420 nK and about 5 resonances per Gauss for a higher temperature of 800 nK.
This observation of these resonances was predicted by the simulations of Ref.~\cite{Petrov2012}, which showed that strong anisotropic
interactions between Dy atoms will lead to the appearance of strong resonances. The theoretical calculations of the Dy 
Feshbach spectrum, however, were performed at a much lower collisional energy of $E/k_B$ = 30 nK and exhibited a lower 
resonance density. In order to give quantitative insight into collisional dynamics of magnetic dysprosium the theory must optimize the model 
parameters  and  conduct calculations at the experimental conditions of Ref.~\cite{BLev2014}.
It seems likely that Dy is more anisotropic than previously thought \cite{Petrov2012}. In addition,  at higher temperatures
around 400-800 nK non-zero partial wave collisions may become important and increase the resonance density.

\subsection{Study of Er atom collisions in a MOT and an optical trap}\label{sec:er}

\begin{figure}
\includegraphics[scale=0.9,angle=0]{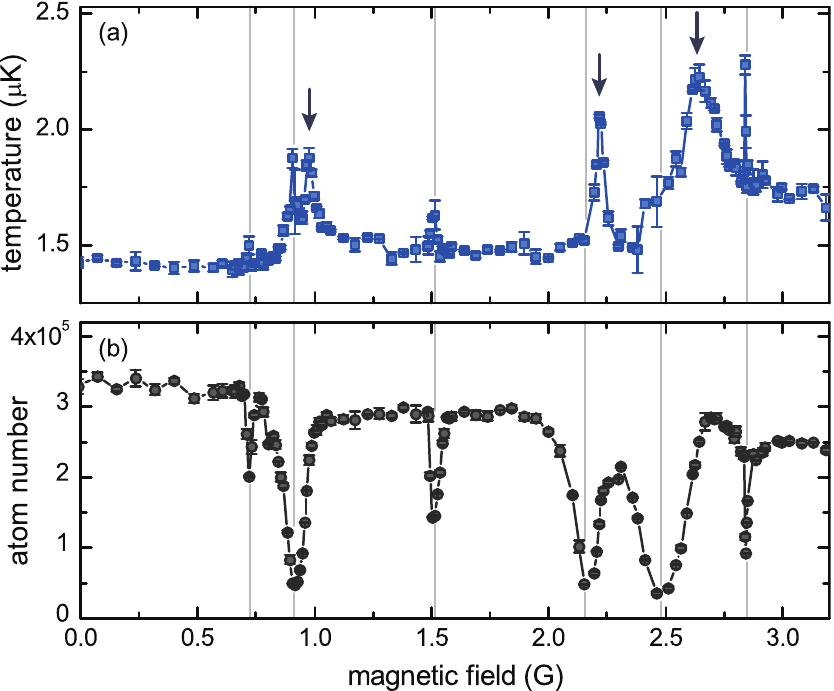}
\caption{Experimental observation of Feshbach resonances in collisions of $m_J=-6$ $^{168}$Er atoms.
The measured temperature (a) and atom number (b) are plotted as a
function of magnetic field. Reproduced with permission of Ref.~\cite{Ferlaino2012B}.
}
\label{fig:FR_Er}
\end{figure}

Exploration of collisional dynamics between the ultracold erbium atoms
began in 2006 by Dr.~McClelland's group at NIST \cite{McClelland}. They
demonstrated the operation of a magneto-optical trap (MOT) and laser cooling
of Er to millikelvin temperatures.  Later the group of Prof.~Ferlaino
at the University of Innsbruck, adapted the cooling
techniques of \cite{McClelland} and applied a narrow-line MOT allowing
them to further cool Er atoms. They reached  temperatures of 15 $\mu$K
\cite{Ferlaino2012A}.  After directly loading  atoms into an optical
dipole trap and using evaporative cooling they created a pure $^{168}$Er
BEC \cite{Ferlaino2012B}.  At magnetic fields below
3 Gauss they immediately observed six Feshbach resonances as shown in
Fig.~\ref{fig:FR_Er}. The presence of low-field resonances in the
Er interactions opens extraordinary possibilities to study dipolar
properties for atoms with a strong coupling between various partial
waves and Zeeman sublevels.

\subsection{Evidence for a strong collisional anisotropy in Er interactions}

A further study by this group using high-resolution trap-loss
spectroscopy~\cite{AFrisch2014} revealed a dense forest of
Feshbach resonances for magnetic fields from 0 G to 70 G.
In fact, they observed for two bosonic isotopes, $^{168}$Er and $^{166}$Er, both in their lowest Zeeman
sublevel $m_j = -6$  this unprecedented large number of
resonant features with a mean density $\bar{\rho}$ of 3 resonances
per Gauss.  An even more dense spectrum  of 26 resonances per Gauss
was obtained for an optically-trapped sample of fermionic $^{167}$Er
atoms in their lowest Zeeman sublevel, $m_F = -19/2$.  

Based on the enormous number of resonances the experimentalists with
support of the theorists set up to investigate the statistical properties
of the observed spectra for the bosonic atoms. First, the authors of ~\cite{AFrisch2014}
performed an analysis of the nearest-neighbor spacings (NNS) and interpreted
their results using distributions derived in random matrix theory (RMT), originally introduced
by \cite{Wigner1951,Dyson1963}. RMT attempts to characterize the presence or absence of correlations between
levels or in this case Feshbach resonance positions. In parallel, several first-principle
coupled-channel (cc) calculations of Er Feshbach spectra were performed.
A resonance spectrum of the scattering length was obtained by only  including 
channels with even partial waves $\ell$ from 0 up to $L_{\rm max}$.
Calculations were performed up to $L_{\rm max}=20$.  The theoretical simulations
also found a large number of resonances between $B=0$ G and 70 G, which could also by analyzed in terms of
resonance distributions compared to predictions of RMT.

\begin{figure}
\includegraphics[width=0.48\textwidth,trim=0 120 0 0,clip]{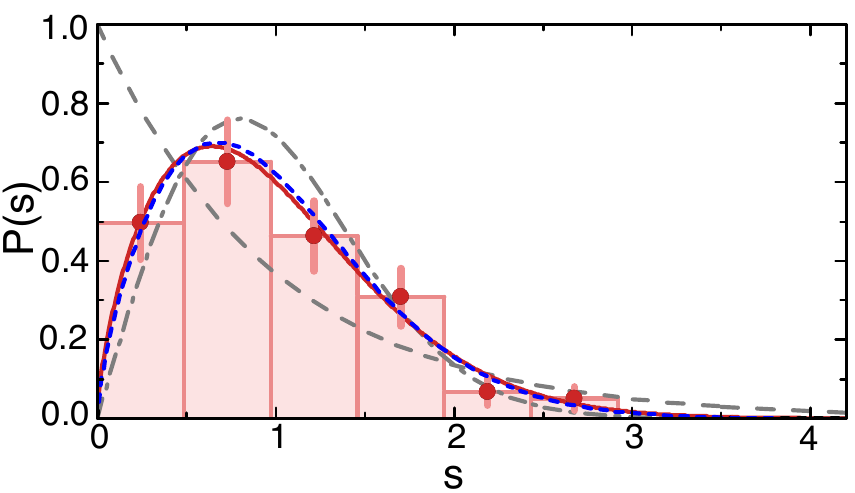}
\caption{The normalized distribution $P(s)$ of nearest-neighbor spacings (NNS) of $^{168}$Er Feshbach resonances
as a function of dimensionless $s=\Delta B{\bar \rho}$, where the $\Delta B$ are NNS and $\bar \rho$ is the mean resonance density per unit field strength. 
The experimental data is shown as a bar graph and  filled red circles with error bars.
The dashed gray, dash-dotted gray, and solid red curves are Poisson, Wigner-Dyson, and Brody distributions fit to the experimental data, respectively.
The dotted blue line is a Brody distribution fit to the distribution of NNS of a
close-coupling calculation where partial waves up to $L_{\rm max}=20$ have been included.
Reproduced with permission of Ref.~\cite{AFrisch2014}.}
\label{fig:statistics}
\end{figure}

\subsection{Statistical description of a Feshbach spectrum}

Figure~\ref{fig:statistics} shows the distribution of the nearest-neighbor spacings (NNS)
between Feshbach resonances for the $^{168}$Er isotope for fields between
$B$=30 G to 70 G and grouping resonance spacings $\Delta B$ in bins with a width of 160 mG.
These spacings scaled to the mean spacing, $s$, were then fit to a Poisson distribution $P(s)=\exp(-s)$
for non-interaction levels,  the Wigner-Dyson distribution $P(s)= (\pi/2)s \exp(-\pi s^2/4) $  characterizing strongly
interacting levels, and the Brody distribution, which is a one-parameter function that smoothly connects
between the Poisson and Wigner-Dyson distribution. The authors of ~\cite{AFrisch2014}
concluded that the NNS of erbium closely resembled a Wigner-Dyson distribution. 

In addition, Fig.~\ref{fig:statistics}  shows the Brody distribution fit to the NNS of the theoretical coupled-channel spectrum  with $L_{\rm max}=20$ and
using a collision energy of $E/k_B=360$ nK is shown. The agreement between the experimental and theoretical Brody distributions is impressive. 
This is the more surprising as it should be noted
that for $L_{\rm max}=20$ the coupled-channels calculation did not yet converge. For example
the resonance density was still increasing with increasing $L_{\rm max}$ and for $L_{\rm max}=20$ the mean density was only ${\bar\rho}=1.5$ per Gauss.

\begin{figure}
\includegraphics[scale=0.95,angle=0]{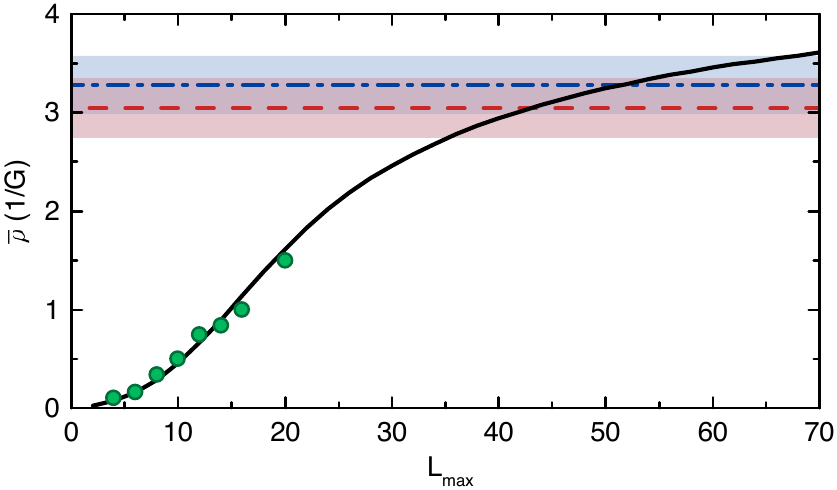}
\caption{Mean resonance density $\bar\rho$ for bosonic Er as a function of the largest included partial wave $L_{\rm max}$ of 
theoretical close-coupling calculations. The cc-calculated densities are
shown by green circles up to $L_{\rm max}$ = 20. The solid black line shows an analytical estimate of mean density for $L_{\rm max}$ up to 70. 
The experimentally measured mean resonance densities  are presented for $^{168}$Er (dash line) and for $^{166}$Er (dash-dotted line) 
with one sigma confidence bands (shaded areas). Reproduced with permission of Ref.~\cite{AFrisch2014}.
}
\label{fig:Lmax}
\end{figure}

The behavior of the mean resonance density of Feshbach resonances obtained from cc calculations was further studied by systematically
increasing the largest included partial wave $L_{\rm max}$ and are shown in Fig.~\ref{fig:Lmax}.  
The figure also shows an estimate of ${\bar\rho}$ for larger $L_{\rm max}$
by essentially counting the number of weakly-bound rovibrational states of the closed channels. This simple
approach relied on a separation of energy scales:  For small/large interatomic separations  the anisotropic couplings are large/small compared 
to the Zeeman interaction and rotational or coriolis forces, respectively, and that, crucially, in the radial cross-over region
the isotropic part of the potentials is much deeper than the Zeeman and rotational splittings. 
Reference \cite{AFrisch2014} then concluded that at least 40 partial waves will be needed to explain the experimental resonance density.

\subsection{Dipolar effects on fermionic atom cooling}

We finish this section by noting that Erbium has one stable fermionic isotope, $^{167}$Er, with non-zero nuclear spin.
First experiments for this isotope have only recently been performed. Not surprisingly the magnetic
dipole-dipole interaction between these Er atoms plays a prominent role in their cooling.
Reference~\cite{Aikawa2014} showed that fermionic atoms, all in the same magnetic sublevel, can still be thermalized eventhough such atoms can not
collide by even partial waves (and in particular the $s$-wave) as Fermi statistics forbids such channels. They must thermalize 
by $p$-wave collisions instead.  This thermalization occurred  with cross-sections that are consistent with perturbative predictions from universal dipole
scattering theory \cite{Ticknor2008,Bohn2009}.

\section{Magnetic control of mixed gases of ground and meta-stable rare-earth atoms} \label{sec:meta}

The collisional characteristics of the meta-stable Yb$^*$($^3$P$_2$) atoms were recently explored by Prof.~Takahashi's
group at Kyoto University, Japan, and Prof.~Gupta's group at the University
of Washington \cite{Takahashi2013,Gupta2014}. Again in contrast to the well
studied alkali-metal atom collisions, ultracold collisions between meta-stable
rare-earth atoms are highly anisotropic due to the interplay between  interaction anisotropies
and spin-orbit coupling.  The first experimental realization of optical trapping
of ultracold $^{174}$Yb$^*$($^3$P$_2$) atoms and measurement of its 
inelastic collisional rate was performed by \cite{Takahashi2008}. Fine-structure
changing collisions were suggested to be the main source of these
losses. Later, the same group measured the dynamic polarizability of
$^3$P$_2$ magnetic sublevels at laser wavelengths of 532 nm \cite{Takahashi2010} and 1070 nm \cite{Hara2013}
by performing a high-resolution spectroscopy in a Yb Bose condensate.

Historically, considerable attention has been given to the theory of fine-structure
changing  collisions of atoms  in $P$ electronic states with structureless atoms. Spin-orbit relaxation 
in such systems was found to be very efficient due to the interplay with interaction anisotropies.
The orientation-dependent cross-section for fine-structure transitions in collisions
between  ground state He and  $^2$P Na  atoms near room-temperatures
was first analyzed in Ref.~\cite{Reid1969}. The oscillatory behavior of the cross-section 
with collision energy was shown to be related to shape resonances
in the elastic scattering channels. The theory of fine-structure transitions in collisions between a
proton and a fluorine atom in its ground $^2$P state was formulated in terms of molecular states  
and treated by the quantum close-coupling method in Ref.~\cite{Mies1973}.  

Fine-structure changing collisions of alkaline-metal atoms in a $^3$P state with 
He atoms was studied in Ref.~\cite{Alexander1983}. They reported a significant increase in the 
cross-section due to orientational anisotropies leading to coriolis coupling between the orbital
and electronic angular momenta. In Ref.~\cite{Pirani1988} the interaction of oxygen $^3$P-state atoms 
with the rare gases was studied. The authors  found that the interaction anisotropy increases from He to Xe due to an increasing 
contribution  from the excited ionic states. This investigation emphasized the effect
of anisotropy in the van-der-Waals interaction on collisional dynamics of the open shell atoms.    
Collisions between two $^3$P oxygen atoms was studied in Ref.~\cite{Zygelman1994}. As in previous 
calculations the authors estimate the cross-section of a fine-structure transitions using a full quantum 
close-coupling theory. They found that inclusion of the fine-structure splitting in the model has a 
dramatic effect on the transition cross-section. In 2008 a combined experimental and theoretical study 
\cite{Maxwell2008} of cold 1 K collisions between open $P$-shell bismuth and helium atoms in the presence 
of a magnetic field  demonstrated  strong Zeeman relaxation attributed to the combined effect of 
interaction anisotropy and spin-orbit coupling. 

Finally, we note that anisotropies in atom-molecule van der Waals complexes containing $P$-states systems were studied in 
Refs.~\cite{Hutson1,Hutson2}. It was emphasized that the long-range 
intermolecular forces have a significant influence on the collisional dynamics by orienting
the reactants during the collision. In addition, spin-orbit couplings have a strong effect on
the long-range forces.

\subsection{Homonuclear ground and meta-stable state collisions}

\begin{figure}[t]
\begin{center}
\includegraphics[width=0.4\textwidth,trim=0 15 250 0,clip]{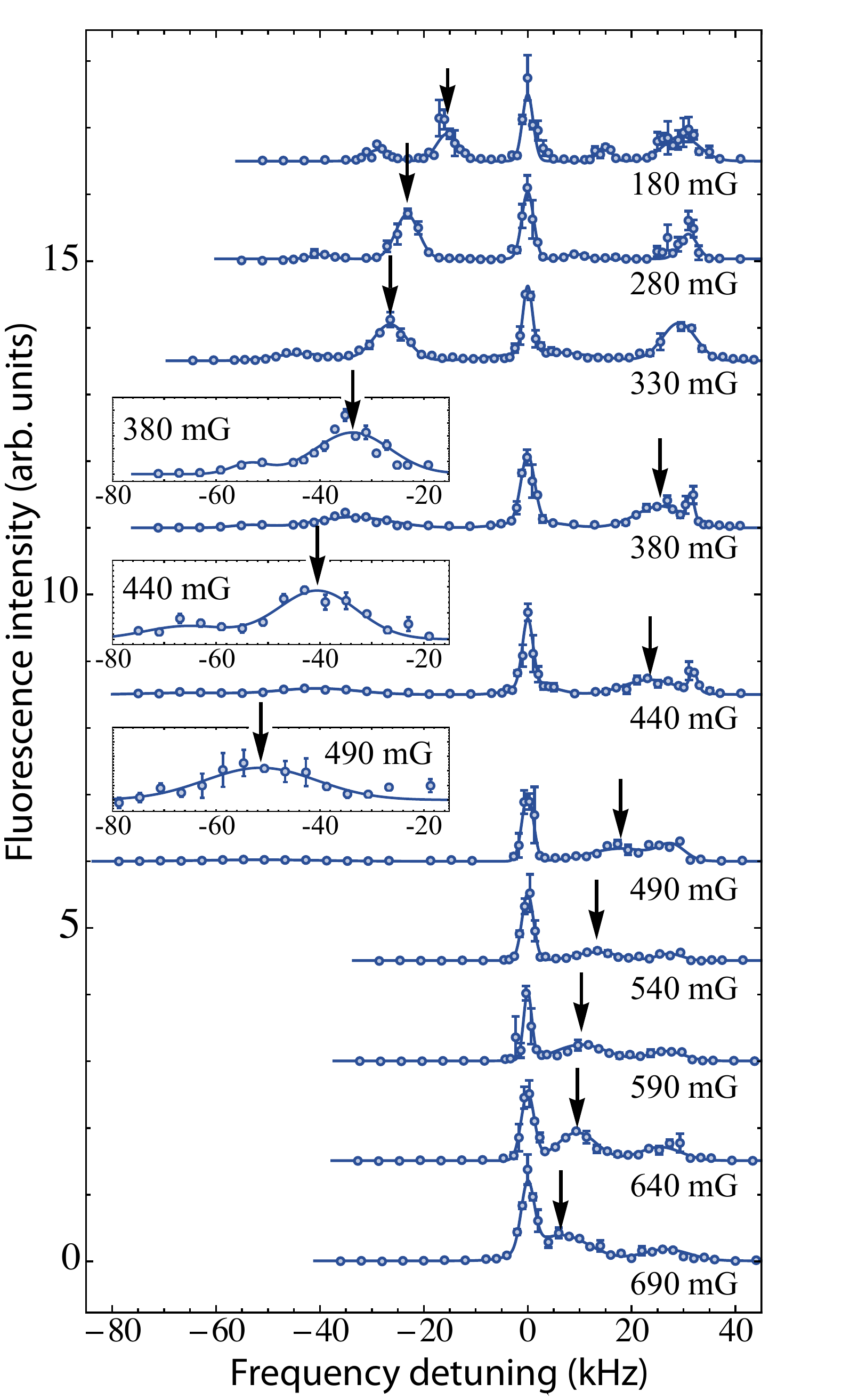}
\end{center}
\caption{Excitation spectra near the $^3$P$_2$ ($m_J$ = +2) state of $^{174}$Yb in an optical lattice at various
magnetic fields below 1 Gauss. Lattice sites contain either one or two atoms. Peaks due to doubly occupied sites are indicated by arrows
and change their location as a function of magnetic field.
Reproduced with permission of Ref.~\cite{Takahashi2013}.
}
\label{YbYbExc}
\end{figure}

Feshbach resonances in ultra cold meta-stable atom collisions is a result of strong interaction 
anisotropies that depend on the orientation of the interatomic axis relative to an external magnetic field.
Resonances of this nature were recently observed by \cite{Takahashi2013}  for homonuclear collisions
between a ground and a meta-stable $^3$P$_2$ $^{174}$Yb atom held in doubly-occupied sites of an optical lattice as shown in
Fig.~\ref{YbYbExc} for magnetic fields below 1 Gauss. 
The spectra in Fig.~\ref{YbYbExc} were obtained by photoassociative spectroscopy near the atomic $^1$S$_0$-$^3$P$_2(m_J=+2)$
transition. The authors infer that a resonance occurs at $B_{\rm res}=360\pm10$ mG.
Reference~\cite{Takahashi2013} also observed a Feshbach resonance in the $^1$S$_0$+$^3$P$_2(m_J=-2)$ collision
of the $^{170}$Yb isotope. Its location is $B_{\rm res}=1.12\pm 0.01$ G.

\subsection{Heteronuclear alkali-metal and metastable rare-earth collisions}

\begin{figure}
\includegraphics[width=0.47\textwidth,trim=0 0 0 0,clip]{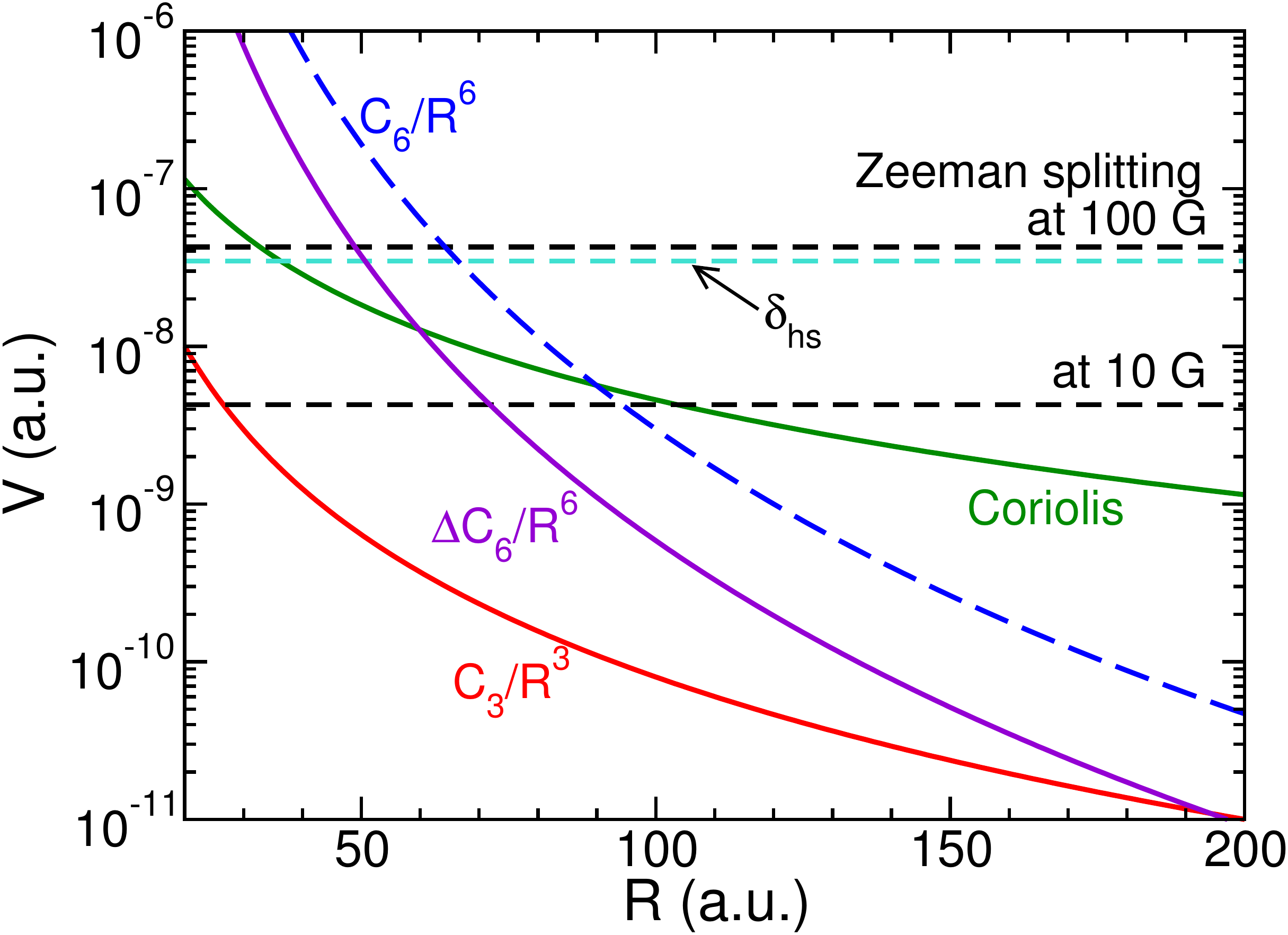}
\caption{Splittings in the Li+Yb$^*(^3$P$_2)$  interaction potentials in atomic units 
as a function of interatomic separation. The Zeeman splitting is $(g_{\rm Li}+g_{\rm Yb^*})\mu_B B$, 
 $\delta_{\rm hs}$ corresponds to the hyperfine splitting of the ground state of Li, 
 the coriolis interaction is $\hbar^2/(2\mu_rR^2)$, the magnetic 
dipole-dipole interaction is $C_3/R^3$, the isotropic and anisotropic dispersion interaction is 
$C_6/R^6$ and $\Delta C_6/R^6$, respectively. Here $g_{\rm Li}$ = 2 and 
$g_{\rm Yb^*}$ = 1.5  are g-factors of Li and Yb$^*(^3$P$_2)$, respectively.
We assumed $C_6$ = 2987.5 a.u. and  $\Delta C_6$ = 585 a.u. \cite{Gupta2014}.}
\label{RelSize}
\end{figure}

In 2011 a quantum-degenerate mixture of fermionic alkali-metal
$^{6}$Li and bosonic $^{174}$Yb atoms was obtained by Dr. Gupta's group \cite{Gupta2011} with the goal to  use photoassociation
create  paramagnetic polar  LiYb molecules.  Now this group has reported on the realization of
an ultracold mixture of $^6$Li and  meta-stable $^{174}$Yb$^*$($^3$P$_2$)  \cite{Gupta2014}.
Measurements
of the two-body inelastic decay coefficients for collisions of the $^3$P$_2$ $m_J = -1$ Zeeman sublevel and a ground state Li atom indicate a low rate 
coefficient of the order of 10$^{-12}$ cm$^{3}$/s. 
As an important aside the authors of Ref.~\cite{Gupta2014} also calculated the dynamic 
polarizability of the ground and meta-stable Yb$^*$ state over a wide range
of laser frequencies allowing the future identification of magic frequencies where both states are identically trapped. 
The long-range dispersion coefficients were also evaluated. Based on these analyses 
Figure~\ref{RelSize} shows the strength of the two major anisotropic long-range interatomic interactions and compares them to Zeeman 
energies and the hyperfine splitting of the Li ground state. When 
the curves for the magnetic dipole or anisotropic dispersion interaction
cross the Zeeman, hyperfine, and/or rotational energies spin flips can occur.
The magnetic dipole-dipole  potential crosses the
Zeeman curves for $B$ = 10 G and 100 G at $R < 50 a_0$, where chemical bonding should play an
important role as well.  

\subsection{Prediction of anisotropy-induced resonances in mixed species collisions}

An important step in the conversion of a weakly-interacting gas  of $^6$Li and
$^{174}$Yb atoms into a strongly interacting one, or even to a gas of  weakly-bound
molecules, is magnetic Feshbach tuning.  
Theoretical work of Ref.~\cite{Brue2012} on ground-state collisions of $^6$Li and
$^{174}$Yb recently predicted that
magnetically tunable Feshbach resonances can exist 
due to a modification of the Li hyperfine coupling in the presence
of the Yb atom.  However, these resonances are expected to be extremely
narrow, on the order of mG, and difficult to observe.  A promising alternative to
observe broader and stronger magnetic Feshbach resonances is to consider interactions between a ground-state Li atom and
a long-lived meta-stable Yb atom.  These meta-stable Feshbach resonances
and its associated weakly-bound meta-stable molecule might be used to
efficiently transfer colliding atoms to a vibrational level of the
absolute molecular ground state.  

Reference~\cite{Hutson2013} performed coupled-channel calculations for $^6$Li+$^{174}$Yb$^*$($^3$P$_2$)  meta-stable collisions 
as a function of magnetic field and showed that  broad and strong magnetic Feshbach resonances can be formed as the
meta-stable Yb$^*$($^3$P$_2$) atom has non-zero electron orbital angular momentum $\vec
L$  and, thus, their interactions are highly
anisotropic. The predicted meta-stable magnetic Feshbach resonances \cite{Hutson2013} may become vulnerable to
decay processes or broadening mechanism that render them
unobservable, such as the spin-orbit interaction of the meta-stable Yb$^*$ atom. 
Collisional resonances of the LiYb$^*$ system could also be broadend by
$R$-dependent spontaneous emission of the interacting atoms. This
process occurs as excited short-lived lithium states  contribute to
the formation of the molecular bond. For the future it would be beneficial to determine
resonance broadening due to this effect.

\section{Conclusion}\label{sec:conclude}

This paper discussed recent advances in our understanding of scattering properties of 
high-spin open-shell atomic systems. In particular, our attention was directed towards 
magnetic atoms with the submerged inner shells, such as chromium, dysprosium, and erbium,
in their ground state as well as ytterbium in its meta-stable state.
These atoms, both bosons and fermions, were successfully cooled to quantum degeneracy and 
confined in optical dipole traps or optical lattices, allowing the study of their collisions at the quantum level.
 
Similar to alkali-metal atom collisions, magnetic Feshbach resonances
represent a powerful tool to control the interactions between these exotic
magnetic atoms. However, the nature and distribution of resonances
for these two systems are completely different. Feshbach resonances in
$s$-wave collisions of magnetic atoms are anisotropy-induced due to a coupling
to rotating molecular states containing  energetically-higher Zeeman sublevels. The long-range
magnetic dipole-dipole and electrostatic interactions are the main
source of this anisotropy. This is in a stark contrast to the alkali-metals,
where the strongest resonances are hyperfine-induced at short range and due to resonant bound states that do not rotate.

In this review we also emphasized a new way of looking at and describing interactions between atoms
and molecules with a multiplicity of internal states. This approach involved the statistical analyses of Feshbach
resonance distribution using random matrix theory. 
The analysis of the resonance distribution of experimentally-observed Feshbach spectra as well as theoretical spectra from quantum-mechanical scattering
calculations in Erbium could be satisfactorily classified in terms of random matrix theory.

Finally, we speculate that there is a growing interest towards the creation of samples of
trapped ultracold mixtures of highly magnetic lanthanide and alkali-metal atoms. These mixtures may have
lower collisional anisotropy and produce less complex resonance spectra possibly leading to
a more easily controllable system.

\section*{Acknowledgements}

The author thanks Dr. E. Tiesinga, Dr. A. Petrov, and Dr. F. Ferlaino for the fruitful discussions
of the various topics of this review. We acknowledge support by the Air-Force Office of Scientific Research and
the National Science Foundation of the USA.


\begin{center} {\Large\bf Bibliography} \end{center}

\bibliographystyle{ieeetr}
\bibliography{BibTexLibraryKotochigova}

\end{document}